\documentclass[12pt,preprint]{aastex}

\usepackage{amssymb}
\usepackage{amsmath}
\usepackage{fixltx2e}
\usepackage{float}
\usepackage{natbib}
\usepackage{color} 

\newcommand{\gtsimeq}{\raisebox{-0.6ex}{$\, \stackrel{\raisebox{-.2ex}%
{$\textstyle >$}}{\sim}\,$}}
\newcommand{\ltsimeq}{\raisebox{-0.6ex}{$\, \stackrel{\raisebox{-.2ex}%
{$\textstyle <$}}{\sim}\,$}}

\shorttitle{Photoevaporation of the Solar Nebula and Jupiter's Abundances}
\shortauthors{Monga and Desch}

\begin{document}

\title{External Photoevaporation of the Solar Nebula: \linebreak Jupiter's Noble Gas Enrichments} 

\author{Nikhil Monga and Steven Desch}
\affil{School of Earth and Space Exploration, Arizona State University, PO Box 871404, Tempe, AZ 85287-1404}

\begin{abstract}
We present a model explaining elemental enrichments in Jupiter's atmosphere, particularly the 
noble gases Ar, Kr, and Xe. 
While He, Ne and O are depleted, seven other elements show similar enrichments ($\sim$3 times solar, 
relative to H). 
Being volatile, Ar is difficult to fractionate from ${\rm H}_{2}$. 
We argue that external photoevaporation by far ultraviolet (FUV) radiation from nearby massive stars
removed ${\rm H}_{2}$, He, and Ne from the solar nebula, but Ar and other species were retained because 
photoevaporation occurred at large heliocentric distances where temperatures were cold enough ($\ltsimeq 30$ K) 
to trap them in amorphous water ice. 
As the solar nebula lost H it became relatively and uniformly enriched in other species.
Our model improves on the similar model of Guillot \& Hueso (2006). 
We recognize that cold temperatures alone do not trap volatiles; continuous water vapor production also 
is necessary. 
We demonstrate that FUV fluxes that photoevaporated the disk generated sufficient water vapor, in regions
$\ltsimeq 30$ K, to trap gas-phase species in amorphous water ice, in solar proportions.
We find more efficient chemical fractionation in the outer disk:
whereas the model of Guillot \& Hueso (2006) predicts a factor of 3 enrichment when only $< 2\%$ of the 
disk mass remains, we find the same enrichments when 30\% of the disk mass remains.
Finally, we predict the presence of $\sim 0.1 \, M_{\oplus}$ of water vapor in the outer solar nebula 
and in protoplanetary disks in H {\sc ii} regions.
\end{abstract}



\section{Introduction}

The abundances of elements in Jupiter's atmosphere have been a mystery since they were measured
by the {\it Galileo} mission.
The {\it Galileo} entry probe's Neutral Mass Spectrometer (NMS) directly measured (down to a depth
of 21 bars, or 150 km) the nine elements C, N, O, S, He, Ne, Ar, Kr and Xe (Niemann et al.\ 1998; 
Mahaffy et al.\ 2000; Wong et al.\ 2004), while {\it Galileo}'s Near Infrared Mapping Spectrometer 
remotely observed phosphine to measure P (Irwin et al.\ 1998). 
In Table 1 and Figure 1 we present the nine (molar) ratios [X/H], where X is any of the non-hydrogen 
species, normalized to the same ratio in the protosolar nebula (Asplund et al.\ 2009).
While not true for He, Ne and O, the seven species C, N, S, P, Ar, Kr, and Xe are enriched with 
respect to H, compared to the ratio in the protosolar nebula gas Jupiter accreted.  
The data are consistent with a uniform enrichment $\approx 3$ (weighted average enrichment is
$2.89 \pm 0.19$).
He and Ne appear depleted, an effect attributed to precipitation of denser He droplets following 
demixing of H and He at Mbar pressures in Jupiter's interior (Stevenson \& Salpeter 1977a,b).
Ne is depleted (enrichment factor 0.12), because it preferentially dissolves into the He droplets 
(Roulston \& Stevenson 1995; Wilson \& Militzer 2010).

Jupiter quite possibly formed with an enrichment of O as well; oxygen may be depleted from its
atmosphere by internal processes.
The factor-of-2 depletion of O (enrichment factor 0.46) has been attributed to the {\it Galileo} 
probe's entry through a $5 \, \mu{\rm m}$ hotspot, a dry region of downwelling gas that may have had 
water removed through meteorological processes (Atreya et al.\ 1999). 
Alternatively, we have suggested that O may similarly be sequestered in the He/Ne droplets at depth 
(Young et al.\ 2014).
A plausible hypothesis is that Jupiter formed with solar abundances of He and Ne and a factor-of-3 
enrichment of O, then saw Ne and O sequestered in He droplets by similar factors $\approx 0.12$, 
resulting in a final enrichment of O of $0.12 \times 2.89 = 0.35$; this is not too different from
the observed enrichment factor 0.48 (Table 1). 
These hypotheses will be tested by the {\it Juno} mission, which will globally measure O abundances,
to depths $> 100$ bar, starting in 2016 (Bolton et al.\ 2010). 

The enrichments of C, N, P, S, Ar, Kr, Xe, and possibly O, are surprising for two reasons.
First, the enrichments are remarkably uniform, considering the quite distinct cosmochemical behaviors
of the different elements.
Second, the enrichments include extremely volatile noble gases like Ar; it is very difficult to
chemically fractionate H and Ar except at very cold temperatures, $\ltsimeq 30$ K.

%
\begin{table}[hb]
\centering
\caption{Elemental enrichments in Jupiter's atmosphere}
\begin{tabular}[b]{crrrrcr}
X & $(X/H)_{\rm Solar Nebula}^{1}$ & Unc. & $(X/H)_{\rm Jupiter}$   & Unc. & Ref. & Enrichment \\
\hline
He & $9.55 \times 10^{-2}$  & $\pm   2\%$ & $7.85 \times 10^{-2}$   & $\pm  2\%$ & 2 & $0.822 \pm 0.025$ \\
Ne & $9.33 \times 10^{-5}$  & $\pm  26\%$ & $1.15 \times 10^{-5}$   & $\pm 19\%$ & 3 & $0.12 \pm 0.04$ \\
Ar & $2.75 \times 10^{-6}$  & $\pm  35\%$ & $9.10 \times 10^{-6}$   & $\pm 19\%$ & 3 & $3.30 \pm 1.32$ \\
Kr & $1.78 \times 10^{-9}$  & $\pm  15\%$ & $4.65 \times 10^{-9}$   & $\pm 19\%$ & 3 & $2.61 \pm 0.63$ \\
Xe & $1.74 \times 10^{-10}$ & $\pm  15\%$ & $4.45 \times 10^{-10}$  & $\pm 19\%$ & 3 & $2.56 \pm 0.62$ \\
C  & $2.95 \times 10^{-4}$  & $\pm  12\%$ & $1.19 \times 10^{-3}$   & $\pm 24\%$ & 4 & $4.03 \pm 1.08$ \\
N  & $7.41 \times 10^{-5}$  & $\pm  12\%$ & $3.32 \times 10^{-4}$   & $\pm 38\%$ & 4 & $4.48 \pm 1.79$ \\
O  & $5.37 \times 10^{-4}$  & $\pm  12\%$ & $2.45 \times 10^{-4}$   & $\pm 33\%$ & 4 & $0.46 \pm 0.16$ \\
S  & $1.45 \times 10^{-5}$  & $\pm   7\%$ & $4.45 \times 10^{-5}$   & $\pm 24\%$ & 4 & $3.08 \pm 0.77$ \\
P  & $2.57 \times 10^{-7}$  & $\pm   7\%$ & $7.7  \times 10^{-7}$   & $\pm  5\%$ & 5 & $3.00 \pm 0.23$ \\
\hline
\multicolumn{7}{l}{1. Asplund et al.\ (2009); 2. Niemann et al.\ 1998; 3. Mahaffy et al.\ 2000} \\
\multicolumn{7}{l}{4. Wong et al.\ 2004; 5. Irwin et al.\ (1998)}
\end{tabular}
\end{table}

%
%
\begin{figure}[h]
\begin{center} 
\includegraphics[width=0.50\paperwidth]{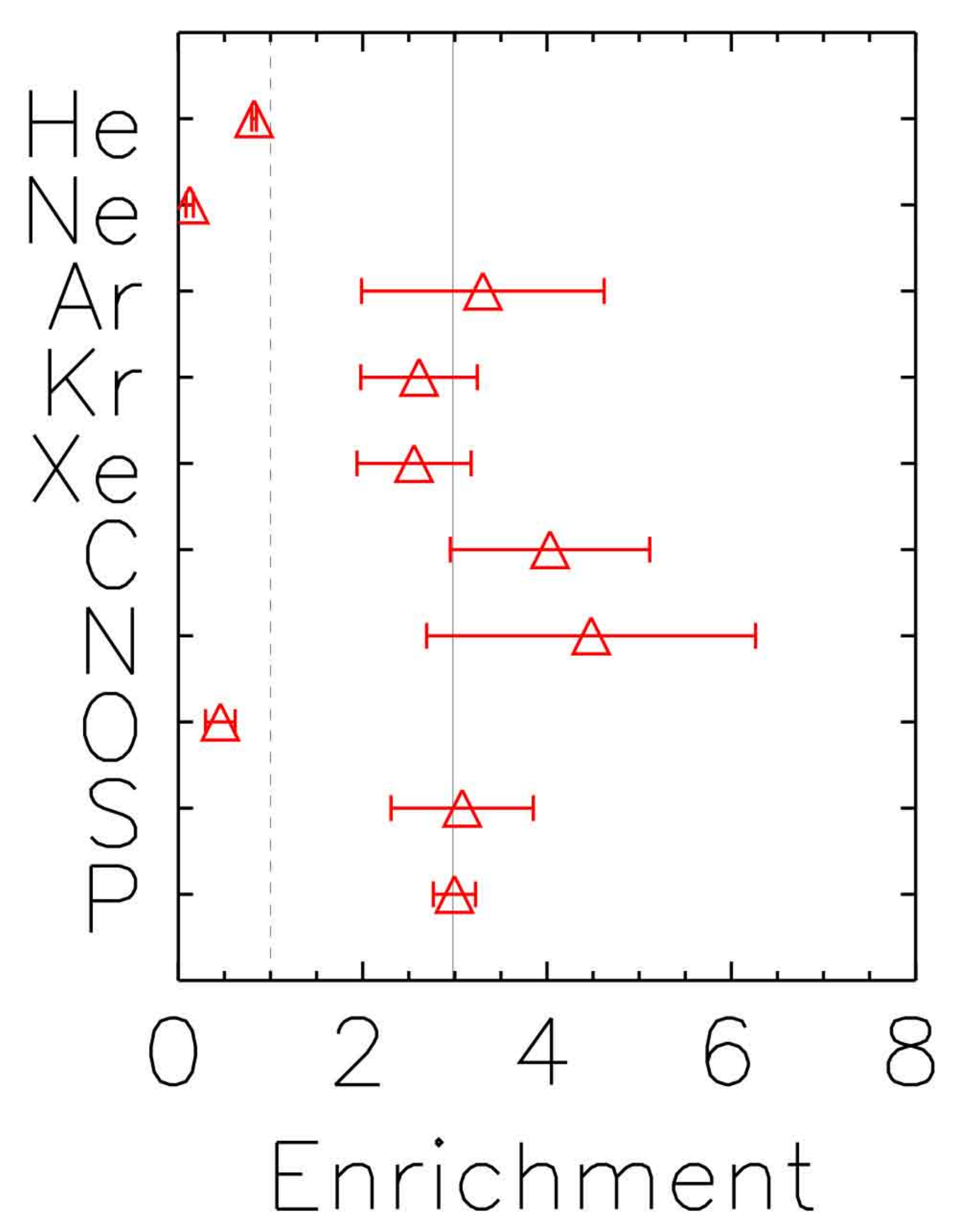}
\caption{Molar abundances of elements relative to H, normalized to abundances in the protosolar
nebula (Asplund et al.\ 2009), as measured by the {\it Galileo} mission (Niemann et al.\ 1998; 
Mahaffy et al.\ 2000; Wong et al.\ 2004). Except for He, Ne and O, all species are enriched in 
Jupiter's atmosphere. The data are consistent with a uniform enrichment $\approx 3$.  The weighted 
average of the enrichments is $2.89 \pm 0.19$ (dashed line).}
\end{center}
\end{figure}
%

Several models have been proposed to explain the mysterious enrichments of noble gases and
other elements in Jupiter's atmosphere. 
It is possible Jupiter accreted ices that were uniformly enriched in all but the most volatile
species.
If the vast majority of the ices Jupiter accreted were amorphous ices formed in a region 
with temperature $\ltsimeq 35$ K, these ices could trap all species (except H, He,
and Ne) equally efficiently (Bar-Nun et al.\ 1985, 1987, 1988; Bar-Nun \& Kleinfeld 1989;
Kouchi 1990; Hudson \& Donn 1991; Owen et al.\ 1992; Givan et al.\ 1996). 
In the ``solar composition icy planetesimal" (SCIP) model of Owen et al.\ (1999) and Owen \&
Encrenaz (2003, 2006), the observed factor-of-3 overabundances of noble gases can be explained
if Jupiter accreted $\sim 10 M_{\oplus}$ of SCIP material.  
A criticism of the SCIP model is that ices in present-day comets appear to be depleted in Ar 
and ${\rm N}_{2}$, and not carrying solar compositions (Cochran et al.\ 2000, 2002; 
Weaver et al.\ 2002). 
It is also the case that Jupiter's formation environment at 5 AU was considerably warmer than
35 K, closer to 55 K; temperatures $< 35$ K would not be reached inside about 18 AU from the Sun
(Chiang \& Goldreich 1997; Lesniak \& Desch 2011).
Jupiter would be expected to accrete considerably more ice from its local environment, depleted
in Ar and ${\rm N}_{2}$, than ice from beyond 18 AU; it would not be expected to exhibit the 
uniform enrichments in noble gases and other elements.
A variation on this model involves trapping species in clathrate hydrates (Gautier et al.\ 2001; 
Hersant et al.\ 2004).  Trapping Ar still requires temperatures $\ltsimeq 35$ K, however, and
clathrates require more O (as water) to be accreted by Jupiter than the model involving 
amorphous ice (Gautier \& Hersant 2005). 

A second model for Jupiter's enrichments, proposed by Guillot \& Hueso (2006; hereafter GH06),
is that Jupiter accreted gas that was solar in composition except for depletions in just the
species H, He and Ne.  
GH06 attributed the depletion of H, He and Ne to loss of gas by external far 
ultraviolet (FUV) photoevaporation, from a region of the disk where all the other species have
condensed into ice grains, so they are not swept up in the photoevaporative flow.  
This model has the attractive feature of predicting uniform enrichments (relative to H) of all 
species less volatile than Ar, i.e., everything except He and Ne. 

Nevertheless, the GH06 model is flawed in several aspects.
GH06 write an equation for the evolution of the enrichment factor of a trace species, ${\cal E}$.
(For example, ${\cal E}_{\rm Ar} = (\Sigma_{\rm Ar} / \Sigma_{\rm H}) / f_{\rm Ar}$, where 
$\Sigma_{\rm Ar}$ and $\Sigma_{\rm H}$ are the column densities of Ar and H at some radius, and 
$f_{\rm Ar}$ is the initial mass fraction of Ar in the protosolar nebula.) 
The diffusion equation written by GH06 for the enrichment factor ${\cal E}$ is
\begin{equation}
\frac{\partial {\cal E}}{\partial t} = 3\nu \left[ \left( \frac{3}{2} +  2 Q \right) \, 
 \frac{1}{r} \frac{\partial {\cal E}}{\partial r} + \frac{\partial^2 {\cal E}}{\partial r^2} \right],
\end{equation}
where $\Sigma$ is the column density of gas, $\nu$ the turbulent viscosity,  
$Q \equiv \partial \ln (\Sigma \nu) / \partial \ln r$, and source/sink terms are nor written here. 
In the limit of small spatial scales, it is seen that turbulent viscosity is overestimated by a 
factor of 3 using this equation.  This equation was derived by incorrectly assuming that trace 
species follow the same diffusion equation as the main gas. 
The correct equation to use, based on Fick's law, is that of Gail (2001):
\begin{equation}
\frac{\partial {\cal E}}{\partial t} = \nu \left[ \left( \frac{5}{2} +  4 Q \right) \, 
 \frac{1}{r} \frac{\partial {\cal E}}{\partial r} + \frac{\partial^2 {\cal E}}{\partial r^2} \right],
\end{equation}
GH06 apparently overestimate the diffusion rate of trace species in the gas.

Another problem concerns the rate at which gas escapes from each annulus (at radius $r$) due to 
photoevaporation.
GH06 use a formula based on the rates calculated by Adams et al.\ (2004), but differing from theirs
(and, more clearly, Equation 13 of Anderson et al.\ 2013) by a factor of $(R_{\rm g} / r)^2$,
where $R_{\rm g}$ is essentially the radius of the sonic point of the photoevaporative flow.
Integration over $2\pi r \, dr$ of Equation 9 of GH06 shows that it does not conform to that of 
Adams et al.\ (2004). 
The GH06 model overestimates the mass loss from the top surface.
It ignores the considerable outward radial transport that occurs in photoevaporated disks (Desch 2007), 
which would bring solar-composition gas into the 5 AU region.
It also overestimates the ability of enriched gas to 
diffuse back into the inner disk. 

In addition, GH06 assume that noble gases and other species are trapped in small icy grains that 
would be lost from the gas if caught in the photoevaporative flow.
They invoke settling of dust to the disk midplane to explain why grains are not lost from the
top surface of the disk.  
Meanwhile, at the outer edge of the disk, they assume icy grains {\it are} lost, so that mass 
loss at the outer edge does not enrich the nebular gas in Ar or other species.  
These assumptions have the result that their model disks see a factor-of-3 enrichment only after 
they suffer loss of at least 98\% of their mass (Figure 3 of GH06). 
In fact, if the temperatures of the photoevaporated gas are 200 K or less, 
factor-of-3 enrichments are never seen.
As we show below, many icy grains do remain suspended in the gas and yet are not lost in the 
photoevaporative flow, even from the disk outer edge.
The GH06 model also discounts the inward migration of large volatile-bearing ice grains, which we 
argue is important. 
Despite these shortcomings, we favor the basic idea of the GH06 model, that photoevaporation
depletes the disk of H and Jupiter accretes gas relatively enriched in other species.
Below we describe a disk model that retains this idea but also complies with modeled temperatures 
and radial transport in photoevaporated disks.

A more fundamental problem of the GH06 model, the SCIP model, and indeed any model seeking to 
fractionate Ar and ${\rm H}_{2}$, is the issue of how volatiles are trapped in ice at low temperatures. 
Cold temperatures $\ltsimeq 35$ K are necessary but not sufficient conditions to trap Ar and
other noble gases in amorphous ice.
The experiments of Bar-Nun et al.\ (1985, 1988) and others revealed that gases like Ar could be 
trapped in amorphous ice {\it as the ice was forming}, via deposition of water vapor onto a
very cold substrate. 
The ice that forms is amorphous precisely because of this vapor deposition, and the trapping
is attributable to the speed at which vapor freezes.
If amorphous ice were not {\it forming}, then Ar vapor could condense only by adsorbing onto
the ice or condensing as Ar ice.
But adsorption can only produce a monolayer, which we show (\S 4.1) is insufficient to trap much Ar. 
Trapping of Ar and other species in amorphous ice, in solar proportions, requires the 
simultaneous conditions of plentiful water vapor, and temperatures $< 35$ K. 
This raises the question of how water vapor is generated in such a cold region. 
Below we describe how FUV radiation---the same FUV radiation responsible for driving
the photoevaporative outflow---will also photodesorb water molecules from ice, generating
vapor even in the coldest parts of the disk.
This same process (though usually involving FUV radiation from the central star) has been invoked 
to explain the observed abundances of very cold water vapor in protoplanetary disks such as TW Hya 
(Hogerheijde et al.\ 2011) and DG Tau (Podio et al.\ 2013), and has been invoked by Ciesla (2014) 
as a mechanism for forming amorphous ice in the outer solar nebula. 

These new insights have enabled us to develop a more complete model of how solar nebula gas
can be enriched in elements (relative to H) over time. 
This model is based on that of GH06 but accounts for several additional physical effects.
The model is illustrated in Figure 2. 
The disk is externally photoevaporated, losing mass at its outer edge at $\approx 50$ AU (Adams et al.\ 2004).
As a result, the net flow of mass is outward through the outer solar system (Desch 2007). 
As ices move through the outer solar nebula they are subject to irradiation by FUV radiation, the same
radiation from nearby massive stars that is causing the disk's photoevaporation.  
The UV flux impinging the disk is a factor $G_0$ times the interstellar radiation field $F_{\rm ISRF}$.
At the disk midplane, where the bulk of ices are, the UV flux is attenuated by a factor $e^{-\tau}$, 
where $\tau$ is the UV optical depth.  
UV photodesorbs water molecules from the ice grains, resulting in a non-zero steady-state 
partial pressure of water vapor.
Due to the cold temperatures $\approx 25$ K in the outer disk near 50 AU, water vapor refreezes as 
amorphous ice when it encounters ice or dust grains.
As it freezes, it also traps whatever other species are in the gas phase, including Ar, Kr, and Xe
(but not ${\rm H}_{2}$, He or Ne). 
Although water vapor is likely to condense first onto the smallest (micron-sized) particles with the 
largest collective surface area, particles quickly grow to sizes large enough ($\gtsimeq 3 \, \mu{\rm m}$)
to dynamically decouple from the gas and avoid being carried out of the disk by the photoevaporative 
outflow. 
Trapped gases, including Ar, are retained by the disk, and only ${\rm H}_{2}$, He and Ne are lost from
the disk.
From there it is likely that the ice particles will grow larger and experience gas drag, spiralling inward 
as a result.
If these amorphous ice grains move into warmer regions of the disk and heat above $\gtsimeq 100$ K, 
they may undergo a spontaneous phase change and release all or some of the trapped volatiles 
(Bar-Nun et al.\ 1985, 1988; Kouchi \& Kuroda 1990).
Whether or not the grains release these trapped volatiles as gas, they do not leave the disk, and the 
average composition of the disk is relatively enriched in trapped species, including Ar.
If Jupiter accretes this material as it builds its atmosphere, its composition will appear similarly
enriched in all trapped species.
If the disk has lost $\approx 70\%$ of its mass over the time Jupiter accretes its atmosphere, the
uniform enrichment in all species except He and Ne is explained.

%
%
\begin{figure}[hb]
\begin{center} 
\includegraphics[width=0.75\paperwidth]{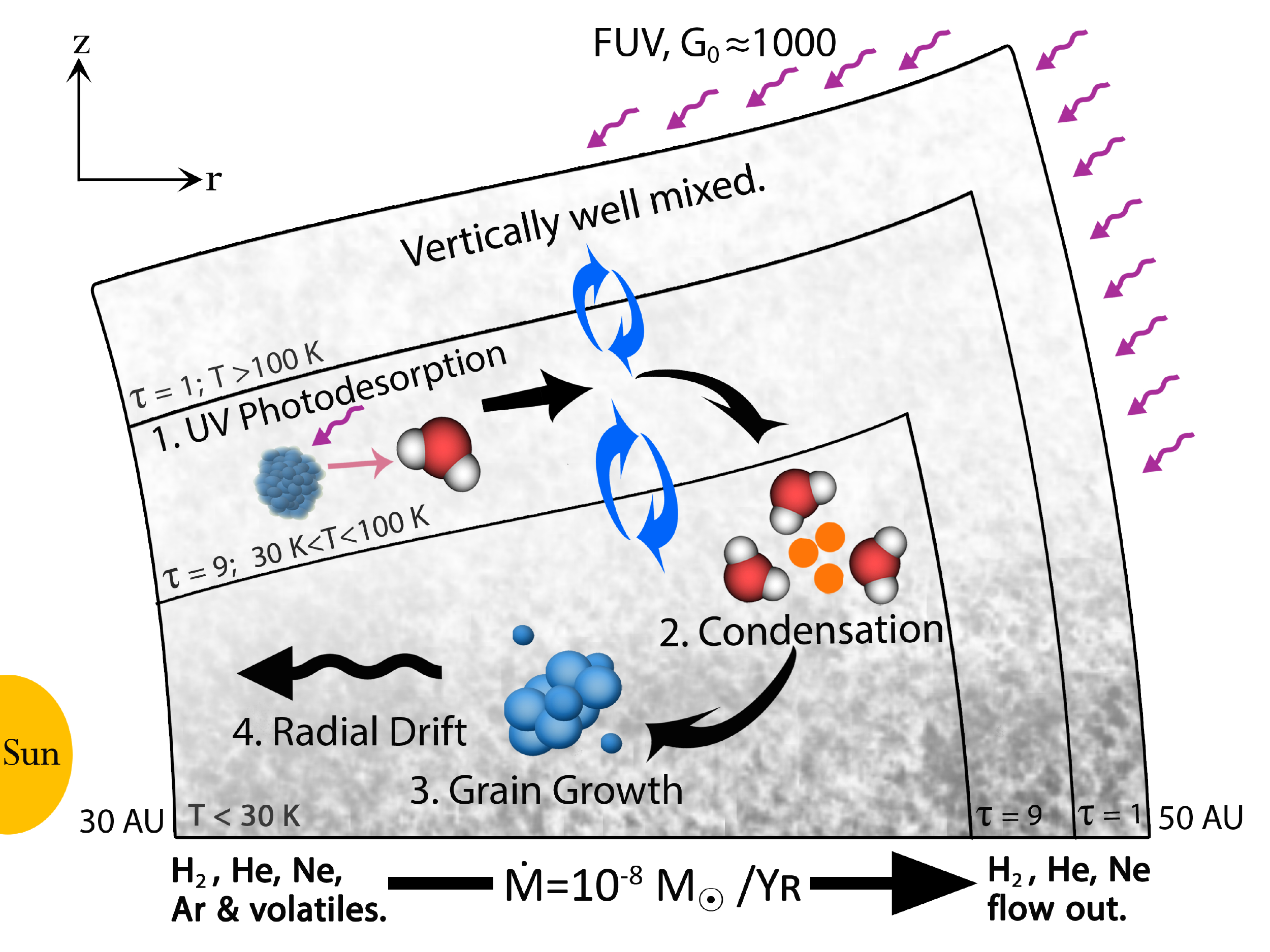}
\caption{Schematic of processes leading to enrichment of elements, Ar in particular, in the solar nebula gas
over time.  Due to external photoevaporation, the net flow of gas is from the inner solar system, outward through 
the outer solar system (at a rate comparable to $\dot{M} = 10^{-8} \, M_{\odot} \, {\rm yr}^{-1}$), to be lost at 
an outer disk edge at $\sim 50$ AU.  {\bf 1.} As ices move through the outer disk they are subject to UV radiation 
that photodesorbs water molecules, generating water vapor.  {\bf 2.} The continuously produced water vapor refreezes 
as amorphous ice, trapping any gas-phase species like Ar as it condenses. The rates are sufficient to remove all species 
except ${\rm H}_{2}$, He and Ne. {\bf 3}. Amorphous ice grains quickly grow to sizes large enough that they dynamically 
decouple from the gas and will not be carried in the photoevaporative flow.  Only ${\rm H}_{2}$, He and Ne are lost 
from the disk's outer edge.  {\bf 4.} As particles grow larger, they drift inward (against the outward flow), there 
they eventually evaporate and release Ar and other species back into the gas phase.  Jupiter accretes solar nebula 
gas that has lost a considerable amount of H and so appears enriched in all other species (except He and Ne).}
\end{center}
\end{figure}

The goal of this paper is to present this newer, more complete model for the enrichments of 
elements in Jupiter's atmosphere and to quantify key physical processes. 
In \S 2 we estimate the likely UV fluxes encountered by protoplanetary disks in high-mass star-forming regions,
and describe the dynamics and evolution of a protoplanetary disk that is experiencing external photoevaporation.
In \S 3 we describe the conditions, including necessary UV flux, that would lead to trapping of noble gases and 
other species in amorphous ice in the outer solar nebula.
In \S 4 we quantify the dynamics of ice particles containing these species.  
We show that in photoevaporated disks like the solar nebula, H, He and Ne are naturally lost from the disk, 
resulting in a gas uniformly enriched in all other species.

\section{Photoevaporation of Protoplanetary Disks}

\subsection{Likely FUV Fluxes}

The model of GH06 invoked external photoevaporation of the disk by the FUV radiation emitted by nearby
massive stars as the mechanism by which the solar nebula lost ${\rm H}_{2}$, He, and Ne. 
It is very common for Sun-like stars to experience intense FUV fluxes from nearby massive O and B stars 
in their birth environments.
{\it Hubble Space Telescope} observations of protoplanetary disks in the Orion Nebula show that 
photoevaporation by FUV is common (Hollenbach et al.\ 1994; Johnstone et al.\ 1998). 
Complete surveys of nearby clusters out to 2 kpc show that 70 to 90\% of all stars are born in dense clusters
(Lada \& Lada 2003), most of which contain O stars that emit $> 10^5 \, L_{\odot}$ of UV radiation
(Adams \& Laughlin 2001).
A preponderance of evidence supports the idea that the Sun was one of the majority of stars born in a 
massive cluster and was exposed to intense FUV radiation (Adams 2010). 
The presence of short-lived radionuclides such as ${}^{26}{\rm Al}$ and ${}^{60}{\rm Fe}$ in the early 
solar system, inferred from meteorites (Wadhwa et al.\ 2007), is evidence for injection of material from
a nearby ($\sim 2$ to 4 pc) supernova into the Sun's disk (Ouellette et al.\ 2007, 2010) or molecular cloud 
(Pan et al.\ 2012).
The progenitor star would be a prodigious source of FUV radiation. 
Isotopically selective UV photodissociation of CO isotopomers in the outer solar nebula is a leading
explanation for the mass-independent oxygen isotope fractionation observed in meteorites (Lyons \& Young 2005).
The classical Kuiper belt has an apparent edge at around 50 AU (Allen et al.\ 2001), which is most easily
explained as en edge due to external photoevaporation (Adams et al.\ 2004; Adams 2010).
The unusual orbit of the Kuiper belt object Sedna requires a close stellar encounter to raise its perihelion 
(Kenyon \& Bromley 2004; Morbidelli \& Levison 2004; Brasser et al.\ 2006). 
For this to be a probable event, the Sun must have been born in a massive cluster. 
All of these lines of evidence support the idea that the Sun's protoplanetary disk experienced signficant 
UV irradiation.

These various lines of evidence can be used to constrain the UV flux, $F_{\rm UV}$, experienced by the solar nebula. 
This is usually expressed as $G_0 = F_{\rm UV} / F_{\rm ISRF}$, where 
$F_{\rm ISRF} = 1.6 \times 10^{-3} \, {\rm erg} \, {\rm cm}^{-2} \, {\rm s}^{-1}$ is the interstellar radiation
field (Habing 1968).
Within local embedded clusters, the probability distribution is strongly peaked to values near $G_0 \approx 3000$,
or more precisely, in the range $G_0 \approx 300$ to $10^4$ (Fatuzzo et al.\ 2008, Adams 2010);
one therefore might expect the FUV flux in the solar nebula to fall in that range.
The value of $G_0$ required to produce the oxygen isotope anomaly depends on the uncertain rate of mixing 
within the disk (parameterized by $\alpha$), but $G_0 \sim 10^3$ to $10^4$ is favored if $\alpha \sim 10^{-2}$
(Lyons et al.\ 2009). 
Likewise, the UV flux needed to produce an edge at 50 AU depends on mixing within the disk; but for 
$\alpha \sim 10^{-2}$ a flux $G_0 \sim 10^3$, corresponding to a mass loss rate 
$\dot{M} \sim 10^{-8} \, M_{\odot} \, {\rm yr}^{-1}$, is required (Adams et al.\ 2004).
Pan et al.\ (2012) invoke a massive ($40 \, M_{\odot}$) star at about  4 pc as the explanation for
the short-lived radionuclides in the solar nebula, which would again result in $G_0 \sim 10^3$. 
The probability of a stellar encounter constrains the stellar density, but this is a remarkably weak
function of the number of stars in the birth cluster (Adams 2010), so cannot be used to constrain the 
FUV flux.  
But the inferred presence of a massive star that went supernova does constrain the total number of stars, $N$,
in the Sun's birth cluster.
Typically a cluster must contain $N \gtsimeq 3000$ stars to have a reasonable chance of hosting a massive
($> 40 \, M_{\odot}$) star that will go supernova in $\approx 4$ Myr (Adams \& Laughlin 2001).
Using the formulas in Adams (2010), this results in $G_0 \approx 1000$ at a distance of 4 pc, the inferred
distance of the Sun from the cluster center based on injection of short-lived radionuclides (Pan et al.\ 2012).
The various lines of evidence support a FUV field with $G_0 \approx 1000$ impinging the Sun's protoplanetary
disk. 

Additional support for the solar nebula being irradiated comes from the inferred structure and 
surface density of the disk. 
Desch (2007), using the starting positions of the giant planets in the `Nice' model (Gomes et al.\ 2005),
reconstructed the so-called ``minimum-mass solar nebula" (MMSN) model of Weidenschilling (1977a).
He derived a surface density that varied with heliocentric distance that matched almost exactly 
the surface density in an steady-state disk being externally photoevaporated, with an outer edge 
$\approx 30$ to 60 AU.
The location of the outer edge depends on the rate of turbulent transport in the disk, but assuming
a turbulent viscosity $\nu = \alpha \, c_{\rm s} H$ (where $c_{\rm s}$ is the sound speed and $H$ the
disk scale height), then $G_0 \approx 10^3$ for $\alpha \approx 10^{-2}$. 

The elevated FUV flux experienced by the solar nebula depleted mass from it.
The FUV photons are absorbed in a thin layer, with optical depth $\tau_{\rm FUV} \approx 1$, at 
the disk's upper surface and outer edge.
Within this thin layer the gas is heated to temperatures $\approx 100$ to 1000 K, which increases 
the gas pressure, driving a thermal wind away from the disk.
Gas loss is possible from all parts of the disk, but gas is predominantly ($\gtsimeq 80\%$) lost from 
where the gravitational pull of the central star is weakest, i.e., at the disk's outer edge
(Adams et al.\ 2004). 

\subsection{Dynamics of Photoevaporated Disks} 

The loss of gas from the disk, especially its outer edge, changes the dynamics of gas at all locations
in the disk. 
Obviously gas will move outward flow through the disk to replenish what is lost at the outer edge.
Desch (2007) showed that the steepness of the surface density profile in the solar nebula demands
this.  He also showed that if a sufficient reservoir of mass existed inside a few AU, then a steady-state
outward is possible that would be uniform between a few AU and the disk edge.  
Subsequent numerical work (Mitchell \& Stewart 2010; Kalyaan et al., in preparation) largely confirms
the result that the outer parts of photoevaporated disks experience outward mass flow that is more
or less uniform and constant. 
From the results of Adams et al.\ (2004; Figure 4), if $G_0 \approx 1000$ and $\alpha$ in the outer disk 
is sufficient to keep the disk edge at $\approx 50$ AU (i.e., $\alpha \sim 10^{-2}$), then that uniform 
outward mass flow $\dot{M}$ is just slightly less than $1 \times 10^{-8} \, M_{\odot} \, {\rm yr}^{-1}$.

There are two important consequences of external photoevaporation on the disk hydrodynamics,
relevant to the issue of volatile trapping. 
First, GH06 argued that inward transport of noble gases must be limited, because otherwise accretion onto 
the star would too rapidly deplete the disk of mass.
In fact, inward radial migration of icy grains and radial diffusion of gas (inward and outward) {\it must} 
occur; but in an externally photoevaporated disk these inward motions, especially of gas trace species, are
frustrated by the outward movement of gas.
Second, GH06 argued for fractionation of ${\rm H}_{2}$ and Ar only during photoevaporative removal of gas 
from the disk's top surface, but escape from the disk's outer edge is dominant. 
Adams et al.\ (2004) found that the total mass lost from the disk's upper and lower surfaces was exceeded
by mass loss from the outer edge by a factor $(G M_{\star} \bar{m} / r_{\rm d} k T)^{1/2}$, 
where $T$ is the temperature to which gas is heated by FUV absorption.
Assuming $M_{\star} = 1 \, M_{\odot}$, $\bar{m} = 1.25 \, m_{\rm p}$, $T \approx 200 \, {\rm K}$, and
$r_{\rm d} = 50 \, {\rm AU}$, this factor is $\approx 4$.
Mass loss from the disk's top surface comprises only $\approx 20\%$ of the photoevaporative 
mass loss from the disk, 80\% of it arising from the outer edge.
(The mass loss onto the star is comparable to the photoevaporative mass loss.)
This led GH06 to greatly overestimate the disk mass loss needed to produce a given enrichment.

\section{Trapping of Volatiles in Amorphous Ice}

\subsection{Production of Water Vapor in Cold Regions of Disks}

Amorphous water ice has the ability, as it condenses, to trap various species in the gas.
Of the species in Jupiter's atmosphere exhibiting enrichments, Ar is the most volatile.
Temperatures $\ltsimeq 25$ to 30 K are required to trap Ar in amorphous ice
(Bar-Nun et al.\ 1985, 1988, 2007; Notesco et al.\ 1996). 
Many protoplanetary disk models show that temperatures this low are reached in the 
solar nebula only at heliocentric distances $> 30$ AU (GH06; Lesniak \& Desch 2011). 
At the expected outer edge of the disk, $r_{\rm d} \approx 55$ AU, midplane temperatures 
are $\approx 25 \, {\rm K}$.
In contrast, where Jupiter forms at 5 AU, temperatures exceed 50 K. 
Trapping of Ar in amorphous ice is not possible where Jupiter forms, but at least at the
disk outer edge there is the opportunity to chemically fractionate Ar and H.

Cold temperatures are a necessary but not sufficient condition for chemical fractionation. 
Trapping also requires that water vapor is in the process of condensing, as the abovementioned
experiments make clear. 
Gas moving outward from the inner disk to the disk outer edge will start with Ar in the 
gas phase but see progressively colder temperatures.  
At some point temperatures will drop below 30 K, but by then all the water ice will have
already condensed. 
We note that the timescale for freezing all water is very short: at temperatures far below
the condensation temperature of water, $\approx 180 \, {\rm K}$ (Lodders 2003), the vapor
pressure of water is negligible, and the sticking coefficient of water molecules is close
to unity. 
In \S 4.1 we show that the timescale to condense a water molecule is $< 10^4$ yr. 
This is much shorter than the time it takes for gas to flow from the inner disk to the 
outer solar nebula where temperatures are low, $\sim 10^6$ yr, which speaks to the need for 
water vapor to be continuously produced.

The need for water vapor to be producd in very cold regions of protoplanetary disks has 
been recognized in a completely different context.
Ceccarelli et al.\ (2005) detected absorption of 464 GHz radiation by very cold 
($< 25 \, {\rm K}$) water (HDO) vapor in the outer regions of the disk around DM Tau.
They and Dominik et al.\ (2005) attributed it to photodesorption by UV or X-ray photons. 
Subsequent searches of DM Tau by Bergin et al.\ (2010) for emission from ${\rm H}_{2}{\rm O}$ vapor,
using the {\it Herschel Space Observatory} {\it Heterodyne Instrument for the Far-Infrared} ({\it HIFI})
instrument, made only one tentative detection.
But more recently, 557 and 1113 GHz emission from cold water vapor in the disk around TW Hya 
was directly observed by Hogerheijde et al.\ (2011), using the same instrument.
The majority of this water emission arises mostly from 100 to 200 AU in TW Hya's disk.
Since both nuclear spin isomers were observed, the ortho-to-para ratio of the water 
vapor was derived, yielding a spin temperature $T_{\rm spin} \approx 13.5 \, {\rm K}$.  
The extremely cold temperatures of the water vapor strongly support models in which water 
is desorbed from icy grains by UV radiation.

\subsection{Trapping of Ar and Volatiles}

Here we demonstrate that the rate of UV photodesorption of water molecules is sufficient to
trap species effectively in amorphous ice, Ar in particular. 
Referring to Figure 2, Ar gas is advected by gas motions, from the inner solar nebula to the
disk outer edge. 
In order for Ar to not be lost from the outer edge of the disk, it must be fully trapped in
amorphous ice during the time it takes to move from 30 AU (where temperatures first drop below
30 K) to 50 AU (the outer edge of the disk).
Ar is trapped by production of water molecules by UV photodesorption in this annulus.
The number of UV molecules impinging the disk is $G_0 F_{\rm ISRF} A / h\nu$,
where $G_0 \sim 10^3$, $A \sim 5000 \, {\rm AU}^2$ is the area of the annulus, and 
$h\nu \approx 10 \, {\rm eV}$ is the average energy of an FUV photon.
The majority of these UV photons are absorbed in the upper surface layers of the disk,
where they heat the disk and drive a photoevaporative outflow, the 20\% of the total
photoevaporative outflow that arises from the disk. 
Any Ar in the gas of these surface layers is lost along with the rest of the gas.
Only after the UV is attentuated by optical depths $\tau \gtsimeq 1$ is the UV flux 
low enough not to significantly heat the gas (Adams et al.\ 2004).
It is at these depths that Ar can be trapped in amorphous ice before it is lost from
the disk.

The total number of water molecules that are generated in this non-escaping region, per time, is
\begin{equation}
\frac{d N_{\rm H2O}}{dt} = \frac{1}{2} \, \frac{ \epsilon G_0 F_{\rm ISRF} A e^{-\tau} }{ h\nu },
\end{equation}

\begin{equation}
 \frac{d N_{\rm H2O}}{dt} \Delta t = \frac{1}{2} \, \frac{ \epsilon G_0 F_{\rm ISRF} A e^{-\tau} t_{cond}}{ h\nu } 
\end{equation}

where $\epsilon \approx 3 \times 10^{-3}$ is the efficiency with which FUV photons
desorb water molecules (Westley et al.\ 1995; \"{O}berg et al.\ 2009).
The factor of $e^{-\tau}$ accounts for the fact that water vapor generated within
one optical depth of the surface will be lost in a photoevaporative flow, and will
not be available to trap Ar in the disk.
The factor of $1/2$ accounts for the fact that water vapor generated at depths has
equal probability of travelling down (toward the midplane) or up (to escape with the flow).
We assume that essentially all UV photons strike dust grains, and that all grains are 
ice-mantled. 
Equation 3 is to be compared to the total number of Ar atoms brought into the region per time, 
which is
\begin{equation}
\frac{d N_{\rm Ar}}{dt} = \frac{ f_{\rm Ar} \dot{M} }{ m_{\rm Ar} },
\end{equation} 
where $m_{\rm Ar} = 36.3$ amu and $f_{\rm Ar} = 1.1 \times 10^{-4}$ are the average mass and
total mass fraction of Ar in a solar composition gas (Lodders 2003), and $\dot{M}$ is the outward
mass flow through the outer disk.
The number of water molecules generated per time in the annulus must exceed the number of 
Ar atoms brought into the annulus per time---a 1:1 ratio of trapped species to ${\rm H}_{2}{\rm O}$ 
molecules is possible in the experiments of Bar-Nun et al.\ (1985, 1988). 
Comparing these two rates we derive the minimum FUV field necessary to lead to trapping of 
all the Ar atoms in the 30-50 AU annulus:
\begin{equation}
G_{0} > \frac{ 2 f_{\rm Ar} \, \dot{M} \, h\nu \, e^{+\tau} }{ m_{\rm Ar} \, \epsilon \, F_{\rm ISRF} \, A}
 = 6.8 \, \left( \frac{ \dot{M} }{ 10^{-8} \, M_{\odot} \, {\rm yr}^{-1} } \right) \, e^{\tau}.
\label{eq:mingo}
\end{equation} 
For the conditions we consider likely in the solar nebula, $\dot{M}$ is slightly less than 
$10^{-8} \, M_{\odot} \, {\rm yr}^{-1}$, and $G_0 \approx 1000$, so using $\tau \approx 1$,
trapping all of the available Ar requires $G_0 \gtsimeq 18$.  
Generation of water vapor by external photoevaporation, sufficient to trap volatiles, is not possible in 
disks that form outside rich clusters, where $G_0 \approx 1$; but 
the expected value for the solar nebula, $G_0 \sim 10^3$, exceeds the critical value by a factor of 56.

The water vapor produced by photodesorption is more than sufficient to trap all of the noble 
gases in their solar proportions, Kr and Xe being much less abundant than Ar.
As the water:gas number ratio drops below a critical value, however, different gas species must compete 
for sites within the amorphous ice (Bar-Nun et al.\ 1988). 
If other species in the gas (not counting ${\rm H}_{2}$, He or Ne) are more abundant
than Ar, they will compete for sites and crowd out Ar and the noble gases.
These competitive effects are not seen (at 30 K) for a water:gas ratio of 1, but we conservatively 
assume they might appear at lower water:gas ratios. 
That is, to trap all the volatiles, there must be more water molecules than atoms/molecules of {\it all} 
the volatiles.
The only other species more abundant than Ar that might exist in the gas phase are C and O, 
as CO, and N, as ${\rm N}_{2}$.
If half the N in a solar composition were in the form of ${\rm N}_{2}$, its mass fraction would
be $4.0 \times 10^{-4}$. 
Assuming the standard molar ratio ${\rm CO}/{\rm H}_{2} = 1 \times 10^{-4}$, the mass fraction
of CO is $1.0 \times 10^{-3}$. 
The water vapor production rate required to trap {\it all} of these species would be a factor 
$\approx 17$ times greater than Equation~\ref{eq:mingo} would suggest, and would require
$G_0 \gtsimeq 320$, still less than the expected value.
It is also worth pointing out that the photodesorption rate is not quite sufficient to sputter all 
of the water molecules trapped in ice.
Using a mass fraction $f_{\rm H2O} = 5.7 \times 10^{-3}$ (Lodders 2003), values $G_0 \gtsimeq 1900$
would be needed to process all the ice entering the annulus.

The total number of UV photons impinging the annulus between 30 and 50 AU is sufficient to 
trap all the Ar, but only if the water vapor is transported from the UV-irradiated region, which
is relatively warm, to the colder regions where $T < 30$ K.
Obseverations (Hogerdeijde et al.\ 2011; Wotike et al.\ 2009)
indicate that most water vapor is generated at a certain height above the disk midplane.
Below this optimal height, the UV that is the cause of photodesorption is too attenuated.
Above this optimal height, fewer ice grains exist, and what water vapor is generated can suffer
photodissociation.
Water vapor generated at optical depths $\tau_{\rm FUV} \ltsimeq 1$ will be lost in a photoevaporative flow,
as mentioned above.

Due to the exponentially increasing density with depth, 
the majority of water vapor will be generated just below this layer
(e.g., Hogerheidje et al.\ 2011).
Half the generated vapor will diffuse upward toward the outflow, and half will move downward to the 
cold disk midplane. 

In \S 4 we demonstrate that the downward moving half eventually condenses
onto cold ($T < 30$ K) icy grains near the disk midplane.
Only then does water vapor condense, and when it does it condenses as amorphous ice capable of 
trapping all the Ar present.  
Note that Ar atoms will rapidly diffuse throughout the vertical column, on timescales
$\sim 1 / (\alpha \Omega)$, where $\alpha$ is the dimensionless constant associated with
turbulence and $\Omega$ is the Keplerian orbital frequency.
Assuming $\alpha \sim 10^{-2}$ yields a vertical mixing timescale $\sim 4000$ yr at 40 AU,
much shorter than the radial transport timescales $\sim 10^6$ yr. 
Even though the condensation of water vapor and trapping of Ar and other species occurs only closer
to the disk midplane, these processes end up depleting the entire vertical column of any gas-phase species.
Although ice grains may be lofted back up to several scale heights and suffer photodesorption,
this process is unable to return a majority of species back to the gas phase, 
because only a small fraction of the mass of ice is photodesorbed.
Any released species are 
readily trapped again at the disk midplane.

\section{Icy Grains in the Outer Solar Nebula} 

\subsection{Mean Free Path of Water Molecules} 

The size distribution and dynamics of icy (or ice-mantled) dust grains in the solar nebula 
in several ways affects the trapping of volatiles in amorphous ice. 
The first important aspect concerns the mean free path of water molecules as they pass through
the gas, which determines how far water vapor can diffuse before it condenses.
To compute this, we assume the dust grains have radii $a$ corresponding to an `MRN-like' distribution, 
so that $dn(a)/da = A \, a^{-3.5}$ between minimum and maximum radii, $a_{\rm min}$ and $a_{\rm max}$,
where $A$ is a constant (Mathis et al.\ 1977).
Based on the sizes of particles found in chondrites, we take $a_{\rm min} = 1 \, \mu{\rm m}$
and $a_{\rm max} = 1 \, {\rm cm}$.
This is not unlike that found in the coagulation calculations of Weidenschilling (1997). 
Using this size distribution, it is straightforward to find the constant $A$ by demanding
that the total mass of dust and ice grains per volume equal some fraction (we take to be 0.01)
of the gas density, $\rho_{\rm g}$:
\begin{equation}
A = \frac{3 \, (0.01) \, \rho_{\rm g}}{8\pi \rho_{\rm s}} \, a_{\rm max}^{-1/2} \,
 \left[ 1 - \left( \frac{a_{\rm min}}{a_{\rm max}} \right)^{1/2} \right]^{-1},
\end{equation}
where $\rho_{\rm s} \approx 2 \, {\rm g} \, {\rm cm}^{-3}$ is the internal density of the 
ice-mantled dust grains.

Before proceeding, we use this result to calculate the mass of Ar that can be trapped in a monolayer 
on icy grain surfaces.
If the number of adsorbed Ar atoms per icy grain surface area is $\phi_{\rm Ar}$, then the mass of
adsorbed Ar, per volume of gas, is easily shown to be 
\begin{equation}
\rho_{\rm Ar,ads} = m_{\rm Ar} \phi_{\rm Ar} \, \int_{a_{\rm min}}^{a_{\rm max}} \, 4\pi a^2 \, \frac{dn}{da} \, da 
 = \frac{ m_{\rm Ar} \phi_{\rm Ar} 3 (0.01) \rho_{\rm g} }{ (a_{\rm min} a_{\rm max})^{1/2} \rho_{\rm s} }.
\end{equation}
The ratio of this mass to the total amount of Ar per volume, $f_{\rm Ar} \rho_{\rm g}$, yields the 
fraction of Ar that can be adsorbed in a monolayer:
\begin{equation}
\frac{ 3 (0.01) m_{\rm Ar} \phi_{\rm Ar} }{ (a_{\rm min} a_{\rm max})^{1/2} \, \rho_{\rm s} f_{\rm Ar} }
 \approx 2 \times 10^{-4},
\end{equation}
where we have generously assumed that each adsorption site is $\approx 2 \, \AA^2$ in area
(Karssemeijer et al.\ 2014), and that Ar atoms occupy 1 out of 20 such sites (competing with other volatiles 
like CO and ${\rm N}_{2}$). 
This justifies the statements made above, that adsorption is insufficient to trap a significant 
fraction of Ar (or CO or ${\rm N}_{2}$). 

Using the value of $A$ derived above, it is then straightforward to find the mean free path $l_{\rm mfp}$ 
of water molecules as they pass through a sea of such grains:
\begin{equation} 
l_{\rm mfp}^{-1} = \int_{a_{\rm min}}^{a_{\rm max}} \, A a^{-3.5} \, \pi a^2 \, da 
= 2\pi A \, a_{\rm min}^{-0.5} \, \left[ 1 - \left( \frac{a_{\rm min}}{a_{\rm max}} \right)^{1/2} \right]
= \frac{ 3 (0.01) \rho_{\rm g} }{ 4\pi \bar{a}^3 } \, \pi \bar{a}^2,
\end{equation} 
where $\bar{a} = (a_{\rm min} a_{\rm max})^{1/2} \approx 100 \, \mu{\rm m}$.
It is as if all of the particles have been replaced with a monodispersion with radius $\approx 100 \, \mu{\rm m}$.
The mean free path can be rewritten in terms of the hydrogen number density, $n_{\rm H}$:
\begin{equation}
l_{\rm mfp} = 760 \, \left( \frac{ n_{\rm H} }{ 10^8 \, {\rm cm}^{-3} } \right)^{-1} \, {\rm AU}.
\label{eq:lmfp} 
\end{equation} 
As soon as a water molecule encounters a dust grain it will condense onto it. 
The lengths traveled by water molecules in the gas before this happens depend on the gas densities
at various heights.

We assume the total disk surface densities from Desch (2007): 
$\Sigma_{\rm tot} \approx 32 \, {\rm g} \, {\rm cm}^{-2}$ at 30 AU, and 
$\Sigma_{\rm tot} \ltsimeq 1 \, {\rm g} \, {\rm cm}^{-2}$ at 50 AU.
We assume temperatures 30 K at 30 AU, and 25 K at 50 AU, so that the scale heights of the disk
are $H \approx 1.8 \, {\rm AU}$ at 30 AU, and $H \approx 3.5 \, {\rm AU}$ at 50 AU. 
Assuming an isothermal disk yields a midplane hydrogen density of 
$n_{\rm H,0} = 2.0 \times 10^{11} \, {\rm cm}^{-3}$ at 30 AU, and 
$n_{\rm H,0} = 3.2 \times 10^{9} \, {\rm cm}^{-3}$ at 50 AU.
At any heliocentric distance, density varies with height $z$ above the midplane
as $n_{\rm H}(z) = n_{\rm H}(0) \, \exp( -z^2 / 2 H^2 )$. 
Comparing with Equation~\ref{eq:lmfp}, the mean free path is small (compared to $H$) if
one uses the gas densities near the midplane; but at a few scale heights above the 
midplane, the mean free path quickly becomes $\gg H$.

The relevant locations at which we must calculate the mean free path are the positions 
of the $\tau_{\rm FUV} = 1$ surface, and the location in the disk where $T < 30 \, {\rm K}$. 
Inspection of Figures 2a and 2b of Adams et al.\ (2004) suggests that while temperatures 
are high ($> 100$ K) in the FUV-irradiated surface layers, they drop below 30 K where 
$A_{\rm V} \gtsimeq 5$, or $\tau_{\rm FUV} \gtsimeq 9$. 
The hydrogen column density above the $\tau_{\rm FUV} = 1$ surface is 
$N_{\rm H} \approx 1 \times 10^{21} \, {\rm cm}^{-2}$ (Adams et al.\ 2004), and that above
the $\tau_{\rm FUV} = 9$, $T = 30 \, {\rm K}$ surface is $N_{\rm H} \approx 9 \times 10^{21} \, {\rm cm}^{-2}$.
These correspond to surface densities $\Sigma = 2 \times 10^{-3} \, {\rm g} \, {\rm cm}^{-2}$ and 
$\Sigma = 2 \times 10^{-2} \, {\rm g} \, {\rm cm}^{-2}$, respectively, showing that the mass of 
gas in these layers is small compared to the mass of the whole disk, even at 50 AU.
The heights $z$ above the midplane of these locations can be found using the approximation
\begin{equation}
N_{\rm H}(> z) \approx \frac{ 0.4 n_{\rm H}(z) H}{ z / H } = 
 \frac{ 0.4 n_{\rm H}(0) \, \exp( -z^2 / 2H^2 ) \, H}{ z / H }
\end{equation}
(Lesniak \& Desch 2011).
At 30 AU, the $\tau_{\rm FUV} = 1$ surface lies at $z = 3.8 H$ above the disk midplane,
and the $T = 30 \, {\rm K}$ surface at about $z = 3.2 H$.
At 50 AU, the $\tau_{\rm FUV} = 1$ surface lies at $z = 2.9 H$ above the disk midplane,
and the $T = 30 \, {\rm K}$ surface at about $z = 2.1 H$.

It is now a simple matter to calculate the mean free paths of water molecules at various heights.
At 30 AU, $l_{\rm mfp} = 520$ AU at the $\tau_{\rm FUV} = 1$ surface, dropping to 
$l_{\rm mfp} = 64$ AU at the $\tau_{\rm FUV} = 9$, $T = 30 \, {\rm K}$ surface,
falling further still to $l_{\rm mfp} = 0.38$ AU at the disk midplane.
At 50 AU, $l_{\rm mfp} = 1590$ AU at the $\tau_{\rm FUV} = 1$ surface, dropping to 
$l_{\rm mfp} = 215$ AU at the $\tau_{\rm FUV} = 9$, $T = 30 \, {\rm K}$ surface,
falling further still to $l_{\rm mfp} = 24$ AU at the disk midplane.
Assuming thermal velocities of water molecules $\approx (8 k T / \pi m_{\rm H2O})^{1/2}$
$\approx 0.2 \, {\rm km} \, {\rm s}^{-1}$, water molecules at the $\tau_{\rm FUV} = 1$
surface at 30 AU would take $1 \times 10^4$ yr to encounter a dust grain and condense;
at 50 AU the timescale is $4 \times 10^4$ yr.
Note that these timescales could be lengthened if even just the smallest dust grains considered
here, with radii $\approx 1 \, \mu{\rm m}$, settle to the midplane.
Using the formulas of Dubrulle et al.\ (1995) we find that micron-sized grains could 
exhibit some small amount of settling. 

The condensation timescale $\sim 10^4$ yr at the $\tau_{\rm FUV}$ surface is actually 
quite long, and in fact water molecules are subject to other processes before they freeze 
out onto a dust grain.
The mean free path of water molecules in the ${\rm H}_{2}$ gas are $< 10^3$ km throughout 
the disk, so water molecules will be strongly coupled to the gas.
Parcels of gas themselves are vertically mixed by turbulence in the disk, especially 
considering that the disk is likely to be fully active to the magnetorotational instability
at these radii (Sano et al.\ 2000), so that $\alpha \sim 10^{-2}$. 
Using a turbulent viscosity $\nu = \alpha H^2 \Omega$, the vertical mixing timescale is 
only $(\Delta Z)^2 / \nu \sim$ $(\Delta Z / H)^2 / (\alpha \Omega)$.
Equating $\Delta Z$ with the distance between the $\tau_{\rm FUV} = 1$ surface and the
$\tau_{\rm FUV} = 9$, $T = 30 \, {\rm K}$ surface, we compute the time for water vapor
generated in the intermediate layer to diffuse downward toward the midplane, to depths
where $T < 30$ K. 
This is on the order of $10^3$ yr: at 30 AU it is 940 yr, and at 50 AU it is 3600 yr. 
Thus water molecules are likely to diffuse to the cold region well before they encounter
a grain and freeze out in the upper layers. 
Once in the lower layers, however, the condensation times are shorter, due to the higher
densities; we find they are 1200 yr at 30 AU and 5400 yr at 50 AU. 
Water vapor is likely to diffuse down into the $T = 30 \, {\rm K}$ layer before it freezes,
but once there it is likely to freeze before diffusing any farther.
These calculations show that most of the water vapor generated in the upper, irradiated
portions of the disk, where temperatures are high ($> 100$ K), actually diffuses down to
much colder regions ($T < 30$ K) before it condenses.

The conclusion that gas diffuses down to colder regions of the disk before condensing is 
fairly robust.  
For example, if the disk is less massive or has evolved, so that the gas densities are 
lower, then ${\rm H}_{2}{\rm O}$ molecules will diffuse even farther downward before freezing.
On the other hand, in many disks dust is observed to have somewhat settled to the midplane.
In their analysis of the spectral energy distributions of Taurus protoplanetary disks, 
D'Alessio et al.\ (2006) found that in many (but not necessarily all) such disks, even small
grains were depleted from the gas layers several scale heights above the midplane, by factors
of 10 or more.
Whether the solar nebula protoplanetary disk experienced similar dust settling is unknown, but
we can estimate its effects. 
Since dust dominates the FUV opacity, the $\tau_{\rm FUV} = 1$ surface can be expected to follow
the dust distribution. 
In disks with dust-depleted upper layers, this surface would be at lower $z$, but the effect is
not large; from Equation 12, if 10 times more column density is required to become optically thick,
then the $\tau_{\rm FUV} = 1$ surface will lie at $z = 3.2 H$ at 30 AU, and $z = 2.1 H$ at 50 AU
(instead of $3.8 H$ and $2.9 H$). 
This means the gas density at the $\tau_{\rm FUV} = 1$ surface increases, by factors $\approx 8$ 
at both radii.
Accordingly the timescales for an ${\rm H}_{2}{\rm O}$ molecule to encounter a grain are reduced
by the same factor, becoming comparable to, but not less than, the timescale for diffusion. 
Thus, for moderate (less than a factor of 10) degrees of dust depletion in the disk upper layers,
our conclusions are unchanged.
Again, it is unknown whether dust was depleted in the upper layers of the outer disk, and in the scenario 
put forth here, in which the gas is turbulent with $\alpha \sim 10^{-3}$, micron-sized dust is not 
expected to settle to the midplane significantly (e.g., Dubrulle et al.\ 1995); but even moderate 
depletions will not change our conclusions.

Further, we can use these numbers to estimate the amount of water vapor in the disk at any time.
The column density of ${\rm H}_{2}{\rm O}$ molecules is 
$N_{\rm H2O} \approx G_0 F_{\rm UV} \, \epsilon \, t_{\rm cond} / (h\nu)$,
where $t_{\rm cond}$ is the effective condensation time.
Based on the above discussion, we assume $t_{\rm cond} = 1000$ yr at 30 AU, and 
$t_{\rm cond} = 4000$ yr at 50 AU. 
We then find $N_{\rm H2O} \sim 1 \times 10^{19} \, {\rm cm}^{-2}$ at 30 AU, and 
$\sim 4 \times 10^{19} \, {\rm cm}^{-2}$ at 50 AU.
The total mass of water vapor between 30 AU and 50 AU  is then found to be $\sim 0.1 \, M_{\oplus}$. 
For comparison, the mass of ice in the same annulus is $\sim 20 \, M_{\oplus}$, so only 
0.5\% of the ice is in vapor form at any moment. 

The column density of Ar atoms is $N_{\rm Ar} = f_{\rm Ar} \Sigma / m_{\rm Ar}$
$\sim 6 \times 10^{19} \, {\rm cm}^{-2}$ at 30 AU, or 
$\sim 2 \times 10^{18} \, {\rm cm}^{-2}$ at 50 AU.
At 30 AU there are slightly more Ar atoms than ${\rm H}_{2}{\rm O}$ molecules in the gas,
possibly leading to less efficient trapping; but that quickly changes to a 1:1 ratio as the 
gas moves past 35 AU.  
Increasingly as gas is radially transported out to 50 AU, the ${\rm H}_{2}{\rm O}:{\rm Ar}$ 
ratio is comfortably larger than unity, allowing efficient trapping of Ar.
Similar trends and conclusions would apply to CO or ${\rm N}_{2}$.

\subsection{Growth of Icy Grains} 

When water vapor condenses onto ice-mantled grains, it is likely to trap Ar and other gases as 
it does, because $T < 30$ K. 
The water vapor is also likely to condense onto the smallest particles, since these have the largest
total surface area.  
For example, in the MRN distribution above, particles with $a < 3 \, \mu{\rm m}$ make up 
only $0.7\%$ of the mass, but 43\% of the surface area. 
Small grains are also the most likely to be carried along with the gas as it enters the photoevaporative
outflow.  
If Ar is trapped in icy grains, but then those icy grains are dynamically coupled to the gas, then
the situation is no different than if Ar had remained in the gas phase.
Only if the Ar can be sequestered into the largest grains, which dynamically decouple from the 
gas, can Ar and ${\rm H}_{2}$ be chemically fractionated.

The size of particle that is retained by the disk can be estimated by examining the forces on it. 
Adams et al.\ (2004) considered the same problem and found that to be retained by the disk even as the 
photoevaporative flow is escaping at speed $V$, the gravitational pull on the grain must exceed the drag 
force on the grain:
\begin{equation}
  \frac{GM_*}{r^2} \left(\frac{4\pi a^3}{3}\rho_s \right) \ge \frac{1}{2}\pi a^2\,\rho_g\,V^2,
\end{equation}
where $r$ is the heliocentric distance. 
The velocity of the outflow is $V \sim 0.1 \, C$, where $C \approx 0.8 \, {\rm km} \, {\rm s}^{-1}$ 
is the sound speed in the FUV-heated gas at the disk edge (assuming $T \approx 200$ K).
Particles therefore must exceed a minimum radius 
\begin{equation}
a \geq \frac{ 3 (1.4 m_{\rm H} n_{\rm H}) V^2 r^2 }{ 8 \rho_{\rm s} G M_{\star} } = 
 0.13 \, \left( \frac{ r }{ 50 \, {\rm AU} } \right)^2 \, 
 \left( \frac{ n_{\rm H} }{ 10^8 \, {\rm cm}^{-3} } \right) \, \mu{\rm m}
\end{equation} 
to be retained by the disk.
[NB: Using the equations of Dubrulle et al.\ (1995), icy grains at 50 AU must exceed radii
$0.6 \, (n_{\rm H} / 10^8 \, {\rm cm}^{-3}) \, \mu{\rm m}$ to settle to the midplane; 
any grains large enough to settle to the midplane, as in the model of GH06, would naturally
{\it not} be lost in the photoevaporative outflow.]
This condition is easily met at the top surface of the disk, where $\tau_{\rm FUV} = 1$ and 
$n_{\rm H} \sim 10^8 \, {\rm cm}^{-3}$. 
At the outer edge of the disk, $n_{\rm H} \approx 3.2 \times 10^{9} \, {\rm cm}^{-3}$, so particles
must be $> 3 \, \mu{\rm m}$ in radius to not be lost from the disk.
As outlined above, small particles $< 3 \, \mu{\rm m}$ in radius have 43\% of the surface area,
and should therefore trap 43\% of the Ar.  
But at least half of the Ar should remain in the disk, not participating in the photoevaporative 
flow.

In fact, the situation is more favorable for retention than this, since icy grains will 
coagulate and shatter and regrow on relatively short timescales. 
In his calculations of these processes, Weidenschilling (1997) found that even at 30 - 50 AU, 
considerable evolution of the solids population takes place in less than a few $\times 10^5$ yr.
This is short compared to the radial transport timescale, which can be derived in two ways.
In the surface density profile of Desch (2007), about $0.01 \, M_{\odot}$ of gas lies between
30 AU and 50 AU.  Assuming a mass flow $\dot{M} \sim 10^{-8} \, M_{\odot} \, {\rm yr}^{-1}$, 
this mass is lost only on timescales $\sim 10^6$ yr.
Or, the timescale for mixing across this region by turbulent diffusion is 
$\sim (20 \, {\rm AU})^2 / \nu$, where $\nu = \alpha H^2 \Omega$.  Using $H = 2.5$ AU and 
$\alpha = 10^{-2}$, we find a radial diffusion timescale $\approx 3 \times 10^5$ yr.
These long radial flow timescales indicate that many small grains will have coagulated into larger 
grains by the time they reach the disk outer edge, and large grains shattered into small grains.
These processes take the Ar originally trapped in small grains and mix it thoroughly into
the ice of all grains.
This allows essentially all of the Ar, or other species, to be trapped in the ice and to
remain in the disk.

Finally, as icy grains grow, they are likely to be subject to significant gas drag, since 
they remain in Keplerian rotation while the pressure-supported gas revolves more slowly.
The headwind they feel causes particles to sprial inward (Weidenschilling 1977b).
The inward velocity is largest for those particles with Stokes number of unity, i.e.,  aerodynamic 
stopping times equal to $\Omega^{-1}$, or particle sizes $a \approx \rho_{\rm g} c_{\rm s} / \rho_{\rm s} \Omega$,
where $c_{\rm s}$ is the local sound speed.
For conditions between 30 and 50 AU, the optimally sized particle radius is $\approx 1$ cm.
As particles coagulate and grow larger, they start to spiral inward faster, reaching the
fastest infall velocity $\approx \rho_{\rm g}^{-1} (dP/dr) / (2\Omega)$ when they reach
centimeter sizes, $P$ being the pressure.
For the conditions we assume, we find this infall velocity is enough to spiral inward 20 AU
in only 1200 years.  
Growth of at least some particles to centimeter size seems assured (Weidenschilling 1997);
these particles, which contain most of the mass of ice and trapped volatiles, will then spiral 
inward. taking the volatiles away from the photoevaporative flow at the disk outer edge, back
into the disk.

\section{Discussion}

The anomalous abundances of species in Jupiter's atmosphere provide key insights into the 
dynamics, evolution, and birthplace of the solar nebula.
Jupiter is enriched, relative to H and relative to the protosolar composition, in the seven
species C, N, S, P, Ar, Kr, and Xe (Niemann et al.\ 1998; Mahaffy et al.\ 2000; Wong et al.\
2004; Irwin et al.\ 1998). 
The data are consistent with a single, uniform enrichment $\approx 2.9$ of these species. 
Jupiter is depleted in He, Ne, and O.
The depletion of He has been attributed to precipitation of He droplets as H and He demix at 
great pressures and depths (Stevenson \& Salpeter 1977).
The depletion of Ne has been attributed to sequestration within these droplets
(Roulston \& Stevenson 1995; Wilson \& Militzer 2010). 
In Young et al.\ (2014) and here we suggest that O may be enriched in Jupiter (by the same
amount as other species, a factor of $\approx 2.9$), but then also depleted from its atmosphere 
by the same global sequestration process that depletes Ne (by a similar factor $\approx 0.12$).
If so, the global enrichment of O would be roughly $0.12 \times 2.9 = 0.35$.
We predict that the {\it Juno} mission will find that O is globally depleted relative to H, 
but not by as much as Ne.
The depletion of O observed by the {\it Galileo} probe, 0.48 times solar, may in fact represent
the global atmospheric O abundance.
Regardless of the case for O, an explanation is still demanded for the apparently uniform 
enrichment in seven species, including the noble gases Kr, Xe, and in particular Ar. 

In broad brush, the model of GH06 explains these uniform enrichments.  
The key aspect of the model is that FUV radiation from nearby massive stars photoevaporates the
disk and removes ${\rm H}_{2}$, He and Ne gas.
If the photoevaporation preferentially occurs where $T \ltsimeq 30 \, {\rm K}$, other species,
including Ar, will be trapped within icy grains too large to escape the disk. 
In this way the disk retains its noble gases and other species as it loses only H, He,
and Ne.  Relative to H, other species appear uniformly enriched. 
In detail, the model of GH06 is incomplete or inconsistent.
As outlined in \S 1, some of the equations used are incorrect.  
The GH06 model also relies on icy grains settling to the midplane to avoid being caught up
in the photoevaporative flow, but in fact any grains larger than micron-sized will not be lost
with the escaping gas. 
By assuming that such grains are in fact lost from the outflow at the disk outer edge, 
GH06 find that significant enrichments of Ar and other species only take place after considerable
mass loss from the disk.
Most importantly, GH06 offer no explanation for how Ar and other species can be trapped in 
icy grains.
They show that temperatures are cold enough for volatiles to be trapped in amorphous ice, but
their model (like others) fails to show how water vapor is produced in this cold region.
Without the condensation of water vapor as amorphous ice, Ar cannot be trapped. 

In this paper we have presented a new model for the enrichment of solar nebula gas. 
We retain the basic idea of the GH06 model while improving it in detail and quantifying some of 
the effects. 
Most importantly, we investigated how FUV irradiation would cause desorption of water molecules
from icy grains in cold regions of the disk, leading to trapping of volatiles. 
We find that the FUV fluxes invoked to cause photoevaporation of the disk, $G_0 \approx 10^3$, 
are more than capable of desorbing enough water vapor to trap all of the Ar, CO, ${\rm N}_{2}$, 
etc.
Solar-composition gas is carried from the inner disk, outward through the 30 - 50 AU region.
Icy grains are irradiated and water molecules desorbed from them. 
The water vapor initially finds itself in a relatively warm ($T \approx 100 - 200$ K) region,
but will diffuse downward to the disk midplane before it condenses onto grains.
Water vapor only condenses where $T \ltsimeq 30$ K, where it condenses as amorphous ice and
can effectively trap volatiles.
At these temperatures, and for these desorption rates, volatiles are fully trapped in their
solar proportions.
Condensation takes timescales of a few $\times 10^3$ yr.  
On longer timescales ($\sim 10^5$ yr), small grains coagulate into large grains and large grains
fragment into small grains (Weidenschilling 1997).  
Trapped volatiles are distributed evenly throughout the grain size distribution.
Gas and particles are advected radially outward on timescales $\sim 0.3 - 1$ Myr.
Icy grains advected to the disk outer edge will not be lost in the photoevaporative flow if 
they are even $3 \, \mu{\rm m}$ in radius; for our assumed size distribution, over 99\% of the
mass of ice avoids escape.  
In addition, on the particle growth timescale $\sim 3 \times 10^5$ yr, icy grains can grow to 
centimeter size, at which point they spiral inward because of gas drag, on very short ($\sim 10^3$ yr)
timescales. 
This inward migration of icy grains with trapped volatiles, as well as radial diffusion, carry
volatiles ``upstream" to smaller heliocentric distances. 

We note that GH06 specifically argue against a scenario in which icy grains radially spiral inward 
and carry trapped volatiles with them, volatiles they then release in the inner disk.
They argue first that different volatiles will be released at different temperatures and therefore
different radii.
Each volatile exists as a gas inside its ``evaporation radius".
Rapid mixing and diffusion lead to a uniform enrichment of each species inside that radius.
The enrichment factor is essentially the entire complement of volatile in the outer disk, divided by
the disk mass inside the evaporation radius.
Because of the different evaporation radii for each volatile, GH06 argue they will be enriched by
unequal amounts, contrary to what is observed.  
This argument ignores the fact that volatiles are not released from amorphous ice at the same
temperatures at which they would condense; volatiles trapped in amorphous ice are released in
discrete temperature changes related to structure changes within the ice that affect the 
specific trapping mechanism (Bar-Nun et al.\ 1985, 1988, 2007; Kouchi \& Kuroda 1990; Hudson \& Donn 1991;
Notesco et al.\ 1996).
Volatiles trapped together are likely to be released together at a similar temperature, and therefore
all trapped species are likely to have similar evaporation radii.
It is worth pointing out that if the transition leading to volatile release is the amorphous-to-crystalline
phase transition at $\approx 143$ K (Kouchi 1990), then possibly the majority of trapped volatiles are only
released as infalling grains reach a heliocentric distance $\approx 1$ AU. 
GH06 also assert that efficient transport of icy grains into the disk's inner region implies a 
rapid loss of these grains onto the star, which then depletes the disk of these grains and the 
noble gases they carry.
This ignores the fact that noble gases will be relased into the gas at or outside a few AU, at which point 
the strong outward flow of gas in photoevaporated disks (Desch 2007) will carry them back into 
the outer disk.  
We do not consider the counterarguments made by GH06 to be pertinent objections.
On the contrary, we argue that forces effectively conspire to keep volatiles from escaping the outer 
disk. 
Large volatile-bearing icy grains migrate inward until they release all their volatiles at a few AU
as the amorphous ice transforms to crystalline.
Gas carrying volatiles moves radially outward through the disk, eventually lost to a photoevaporative
flow at the outer edge, but before it does, noble gases and other volatiles are sequestered in icy grains, 
thanks also to UV photodesorption of water vapor. 

A full calculation of the time- and space-varying enrichments in the outer solar nebula is beyond 
the scope of the present paper, which focuses on presenting the key physical processes and a first
quantification of their effects.
But we can estimate the evolution of the gas-phase Ar enrichment in the outer disk using a simple box model.
We consider the enrichment of Ar in the outer solar nebula, extending from a few AU to 50 AU.  
The mass of gas in this region is $M(t)$, and the mass of Ar in this region is $M_{\rm Ar}(t)$.
The enrichment is ${\cal E}(t) = f_{\rm Ar}^{-1} \, M_{\rm Ar}(t) / M(t)$. 
We assume the enrichment is spatially uniform.  
The radial mixing timescale $t_{\rm mix}$ is not quite short enough to justify this; for example,
$t_{\rm mix} \sim r^2 / \nu$ $\sim (r/H)^2 / (\alpha \Omega)$, which with $\alpha \sim 10^{-2}$ 
and $r/H \sim 17$, yields $t_{\rm mix} \sim 0.7$ Myr at 30 AU.
But inward radial migration of volatile-bearing ices, with subsequent release of volatiles, may
justify this simplifying assumption. 
We assume that gas is lost from the top surface of the disk at a rate $0.2 \dot{M}_{\rm PE}$, and
from the outer edge at a rate $0.8 \dot{M}_{\rm PE}$.  
For now we neglect inflow of gas from the inner disk into this region.
The mass of gas in this annulus is $M(t) = M(0) - \dot{M}_{\rm PE} t$. 
The evolution of the enrichment is given by 
\begin{equation}
\frac{d {\cal E}}{dt} = \frac{d}{dt} \left( \frac{1}{f_{\rm Ar}} \frac{ M_{\rm Ar} }{ M } \right)
                      = \frac{1}{f_{\rm Ar}} \, \frac{1}{M} \frac{dM_{\rm Ar}}{dt} 
                          -{\cal E} \, \frac{1}{M} \, \frac{dM}{dt}.
\end{equation}
Since Ar is entirely in the gas phase at 30 AU, and entirely condensed at 50 AU, we assume on
average it is 50\% condensed in the box region, so that the time rate of change of the Ar mass is
\begin{equation}
\frac{d M_{\rm Ar}}{dt} = -f_{\rm Ar} \frac{{\cal E}}{2} \, (0.8) \dot{M}_{\rm PE}.
\end{equation}
Solving for the enrichment as a function of time, we find
\begin{equation}
{\cal E} = \left( \frac{ M(0) }{ M(t) } \right)^{0.9}.
\end{equation}
This means that the remaining gas is enriched in Ar by a factor of 3 (${\cal E} = 3$)
when the mass in this annulus is 30\% of the original mass. 
We could repeat the calculation, assuming that mass with a solar composition flows into the 
annulus across the 30 AU inner edge, bringing Ar with it. 
If that inflow of mass is a fraction $q$ of the total photoevaporative mass loss $\dot{M}_{\rm PE}$,
then it is straightforward to show that
\begin{equation} 
{\cal E} = -\frac{q}{0.9-q} 
 +\frac{0.9}{0.9-q} \, \left( \frac{ M(0) }{ M(t) } \right)^{(0.9-q)/(1-q)}.
\end{equation}
In this scenario, the 30 to 50 AU annulus outer contains more mass at a given enrichment.
For example, if $q \approx 1/2$, the mass in the outer disk is 45\% of its original value when
the Ar enrichment is ${\cal E} = 3$. 
This simple calculation shows that factor-of-3 enrichments in Ar in the outer solar nebula are
possible while the disk still retains a good fraction of its mass, at least 30\%. 
This is in contrast to the GH06 model, in which factor-of-3 enrichments in the solar nebula are
seen only after the disk has lost $> 98\%$ of its mass (Figure 3 of GH06). 
This is a problem for planet formation models because Saturn, Uranus and Neptune must have taken
considerably longer than Jupiter to grow to the critical mass to capture their ${\rm H}_{2}$/He atmospheres 
(see Desch 2007).
If Jupiter accreted its atmosphere only after 98\% of the disk mass is lost, it is not clear that
any ${\rm H}_{2}$/He gas would remain for the other gas giants to accrete.
Essentially, the disk models presented by GH06 evolve too quickly to be compatible with planet
formation models.

Our disk model presented here is more consistent with planet formation models. 
These tend to show that Jupiter accretes its atmosphere in a very rapid ($\ltsimeq 10^5$ yr) runaway 
process that occurs several Myr after solar system formation (Lissauer et al.\ 2009). 
The composition of Jupiter's atmosphere will basically be a snapshot of the conditions in the solar nebula
at that time. 
To explain Jupiter's noble gas abundances, the disk must have lost about 70\% of its mass at the time
of Jupiter's runaway accretion.  

The time at which only 30\% of the disk mass remains is a function of the external radiation field, 
parameterized through $G_0$, which determines $T$, the temperature of the photoevaporated gas.
It is also a function of the turbulent viscosity in the disk, parameterized using the standard variable, 
$\alpha$.
Assuming an initial disk mass $\approx 0.1 \, M_{\odot}$, $G_0 = 1000$, so that $T \approx 450$ K, and 
assuming $\alpha \sim 10^{-3}$, Anderson et al.\ (2013) find that 70\% of the disk mass is lost after 1.0 Myr,
and after only 0.2 Myr if $\alpha = 10^{-2}$; but these timescales are very sensitive to $T$.
[Notably, Anderson et al.\ (2013) do not calculate mass loss rates by self-consistently calculating $T$,
but rather by interpolating a mass loss rate provided by Adams et al.\ (2004).]
Mitchell \& Stewart (2010) find that 70\% of the disk mass is lost after 1.2 Myr in their canonical 
calculation with $\alpha = 10^{-3}$ and photoevaporated gas temperature $T = 600$ K. 
Increasing $\alpha$ to $10^{-2}$ shortens the disk lifetime by 20\%, but lower values of $T$ greatly 
prolong the disk lifetime; if $T = 100$ K, the disk lifetime is increased by a factor of 3.
Existing calculations suggest that if $\alpha \approx 10^{-3}$ and $T \approx 200$ K, then 70\% mass loss
probably takes $> 2$ Myr.
Calculations of disk evolution in which $\alpha$ is attributable to the magnetorotational instability and
calculated from first principles (Kalyaan et al., in preparation) suggest that while $\alpha \sim 10^{-2}$
in the disk's outermost annuli (and is therefore the value of $\alpha$ relevant to mixing), $\alpha \sim 10^{-3}$ 
is nevertheless the appropriate effective value for determining disk lifetimes, and that 70\% mass loss takes 2 to 3 Myr.

In the models of Lissauer et al.\ (2009), runaway accretion takes place only after $\approx 2.4$ Myr,
but they assume low initial column densities of gas ($700 \, {\rm g} \, {\rm cm}^{-2}$) and solids 
($10 \, {\rm g} \, {\rm cm}^{-2}$) at 5 AU. 
In Desch (2007), Anderson et al.\ (2013), and this work, more massive $\sim 0.1 \, M_{\odot}$ disks
are considered; the models of Desch (2007) entail a gas column density $\approx 1100 \, {\rm g} \, {\rm cm}^{-2}$ 
and a solids column density $\approx 16 \, {\rm g} \, {\rm cm}^{-2}$ at 5 AU.
Based on these densities, we estimate runaway growth could occur more quickly, 

in $< 2$ Myr.
Future work is required to reconcile disk evolution models with models of Jupiter's accretion,
but it seems likely that 70\% mass loss from the disk, and the runaway accretion of Jupiter's 
atmosphere, could easily both occur at a time 
$\approx 2$ Myr 
after solar system formation.

These calculations all assume that Jupiter forms at 5 AU, without significant migration, although
of course Jupiter may have migrated.
Within the context of the Nice model (Tsiganis et al.\ 2005) Jupiter migrates only a few tenths of an
AU, but in other models (Walsh et al.\ 2011; Alexander \& Pascucci 2012) Jupiter migrates several AU.
Significantly, though, these migrations occur after Jupiter has largely formed. 
If Jupiter starts to migrate before it is completely formed, the effect will be to speed up the accretion 
of its atmosphere considerably (Alibert et al.\ 2004). 
In any event, the composition of Jupiter's atmosphere is not sensitive to where Jupiter forms:
radial mixing ensures that throughout the outer solar nebula the gas composition is fairly uniform.

Besides migration, another effect that occurs after Jupiter forms must be considered.
When planets become large enough, they can open low-density gaps in the disk, or even large,
optically thin cavities,  making them ``transition disks".
The median accretion rates of gas onto the host stars are curtailed in transition disks, by 
a factor of $\sim 10$ compared to the median accretion rates in normal protoplanetary disks
(Najita et al.\ 2007; Espaillat et al.\ 2012; Kim et al.\ 2013; see also Rosenfeld et al.\ 2014). 
This implies that if Jupiter opens a gap, then one of the assumptions made above may be incorrect;
namely relatively H-rich gas may not be able to flow into the outer solar system from inside the orbit 
of Jupiter.
The presence of a gap is unlikely to change our conclusions, however.
For one thing, the presence of a gap does not preclude the flow of gas across it.
The above references indicate that mass flows are at least 10\% of the mass flows in disks without gaps,
and direct observations of gas in the gaps also indicates substantial movement of gas through them
(Casassus et al.\ 2013).
For another, the analysis above shows that if gas were not able to flow across the gap, the enrichment of 
outer solar system gas would occur more rapidly.
Reducing the flow of mass into the outer disk is equivalent to decreasing $q$ in Equation 18, but the
results are not sensitive to $q$: factor-of-3 enrichments occur when the gas mass is reduced by factors
of 2 to 3, regardless of $q$. 

If the outer solar nebula disk retained significant gas after a few Myr, as we argue, this would
greatly speed up the growth rates of the planetary cores of Saturn, Uranus, and Neptune (Desch 2007). 
Nevertheless, they must form after Jupiter.
We have argued that the time for the Sun's protoplanetary disk to lose 70\% of its mass was very close 
to the time taken by Jupiter to reach its runaway accretion stage.  
The similarity of these timescales may seem coincidental, but is in fact supported by the observation 
that Saturn, Uranus and Neptune accreted less gas than Jupiter.  
Presumably this is because the disk was substantially depleted by the time they grew large enough to 
initiate runaway accretion. Like GH06, we would expect that the later formation of Saturn, Uranus and Neptune would lead them to
accrete nebular gas more enriched in volatiles than when Jupiter grew. 
Of course, whereas $\approx 90\%$ of Jupiter's mass was accreted in its runaway atmosphere (e.g.,
Lissauer et al.\ 2009), Saturn, Uranus and Neptune are relatively more ice-rich.
More so than for Jupiter, the other gas giants could show two patterns of enrichment, one 
for species accreted in the gas phase (as in this model), and another for species accreted in ices 
(as in the SCIP model of Owen et al.\ 1999). 
Whether the gas giant planets captured noble gases from the nebular gas, or from accreting ices,
we predict that Ar, Kr and Xe should show enrichments relative to H that are similar to each other,
and probably $> 3$.

Finally, one of the most remarkable features of our model is the high mass of water vapor it predicts. 
Our model predicts (\S 4.1) that at any instant, $0.5\%$ of the water ice in the outer solar nebula was 
in the form of water vapor (at least initially, and for our assumed grain size distribution and other
parameters).
Assuming an initial mass between 30 AU and 50 AU of $0.01 \, M_{\odot}$, a solar composition yields
$\approx 17 \, M_{\oplus}$ of water ice, and $\sim 0.1 \, M_{\oplus}$ of water vapor. 
Because the water vapor production rate is proportional to the FUV flux, non-irradiated disks
(with $G_0 = 1$ instead of 1000), we predict, would contain $\sim 10^{-4} \, M_{\oplus}$ of 
photodesorbed water vapor
This is larger than the mass of cold water vapor inferred in the TW Hya disk, $\sim 10^{-6} \, M_{\oplus}$
(Hogerheijde et al.\ 2011), but uncertainties in the condensation timescale, and the lower mass of the
TW Hya disk than the solar nebula, maybe partially explain the discrepancy. 
We maintain that disks in stellar clusters, experiencing intense ($G_0 \sim 10^3$) FUV irradiation,
should experience considerably more photodesorption than TW Hya.
We encourage searches for cold water vapor in protoplanetary disks in H {\sc ii} regions.

\acknowledgments

We gratefully acknowledge support from the NASA Astrobiology Institute.  We thank 
John Shumway and Anusha Kalyaan for helpful discussions.  We thank an anonymous referee
who made helpful suggestions.








\begin{thebibliography}{}
\expandafter\ifx\csname natexlab\endcsname\relax\def\natexlab#1{#1}\fi

\bibitem[{{Adams}(2010)}]{adams2010}
{Adams}, F.~C. 2010, \araa, 48, 47

\bibitem[{{Adams} {et~al.}(2004){Adams}, {Hollenbach}, {Laughlin}, \&
  {Gorti}}]{adams2004}
{Adams}, F.~C., {Hollenbach}, D., {Laughlin}, G., \& {Gorti}, U. 2004, \apj,
  611, 360

\bibitem[{{Adams} \& {Laughlin}(2001)}]{adams2001}
{Adams}, F.~C., \& {Laughlin}, G. 2001, \icarus, 150, 151

\bibitem[{{Alexander} \& {Pascucci}(2012)}]{alexander2012}
{Alexander}, R.~D., \& {Pascucci}, I. 2012, \mnras, 422, L82

\bibitem[{{Alibert} {et~al.}(2004){Alibert}, {Mordasini}, \& {Benz}}]{alibert2004}
{Alibert}, Y., {Mordasini}, C., \& {Benz}, W. 2004, \aa 417, L25. 

\bibitem[{{Allen} {et~al.}(2001){Allen}, {Bernstein}, \&
  {Malhotra}}]{allen2001}
{Allen}, R.~L., {Bernstein}, G.~M., \& {Malhotra}, R. 2001, \apjl, 549, L241

\bibitem[{{Anderson} {et~al.}(2013){Anderson}, {Adams}, \&
  {Calvet}}]{anderson2013}
{Anderson}, K.~R., {Adams}, F.~C., \& {Calvet}, N. 2013, \apj, 774, 9

\bibitem[{{Asplund} {et~al.}(2009){Asplund}, {Grevesse}, {Sauval}, \&
  {Scott}}]{asplund2009}
{Asplund}, M., {Grevesse}, N., {Sauval}, A.~J., \& {Scott}, P. 2009, \araa, 47,
  481

\bibitem[{{Atreya} {et~al.}(1999){Atreya}, {Wong}, {Owen}, {Mahaffy},
  {Niemann}, {de Pater}, {Drossart}, \& {Encrenaz}}]{atreya1999}
{Atreya}, S.~K., {Wong}, M.~H., {Owen}, T.~C., {et~al.} 1999, \planss, 47, 1243

\bibitem[{{Bar-Nun} {et~al.}(1985){Bar-Nun}, {Herman}, {Laufer}, \&
  {Rappaport}}]{barnun1985}
{Bar-Nun}, A., {Herman}, G., {Laufer}, D., \& {Rappaport}, M.~L. 1985, \icarus,
  63, 317

\bibitem[{{Bar-Nun} \& {Kleinfeld}(1989)}]{barnun1989}
{Bar-Nun}, A., \& {Kleinfeld}, I. 1989, \icarus, 80, 243

\bibitem[{{Bar-Nun} {et~al.}(1988){Bar-Nun}, {Kleinfeld}, \&
  {Kochavi}}]{barnun1988}
{Bar-Nun}, A., {Kleinfeld}, I., \& {Kochavi}, E. 1988, \prb, 38, 7749

\bibitem[{{Bar-Nun} {et~al.}(1987){Bar-Nun}, {Prialnik}, {Laufer}, \&
  {Kochavi}}]{barnun1987}
{Bar-Nun}, A., {Prialnik}, D., {Laufer}, D., \& {Kochavi}, E. 1987, Advances in
  Space Research, 7, 45

\bibitem[{{Bergin} {et~al.}(2010){Bergin}, {Hogerheijde}, {Brinch}, {Fogel},
  {Y{\i}ld{\i}z}, {Kristensen}, {van Dishoeck}, {Bell}, {Blake}, {Cernicharo},
  {Dominik}, {Lis}, {Melnick}, {Neufeld}, {Pani{\'c}}, {Pearson}, {Bachiller},
  {Baudry}, {Benedettini}, {Benz}, {Bjerkeli}, {Bontemps}, {Braine},
  {Bruderer}, {Caselli}, {Codella}, {Daniel}, {di Giorgio}, {Doty}, {Encrenaz},
  {Fich}, {Fuente}, {Giannini}, {Goicoechea}, {de Graauw}, {Helmich},
  {Herczeg}, {Herpin}, {Jacq}, {Johnstone}, {J{\o}rgensen}, {Larsson},
  {Liseau}, {Marseille}, {McCoey}, {Nisini}, {Olberg}, {Parise}, {Plume},
  {Risacher}, {Santiago-Garc{\'{\i}}a}, {Saraceno}, {Shipman}, {Tafalla}, {van
  Kempen}, {Visser}, {Wampfler}, {Wyrowski}, {van der Tak}, {Jellema},
  {Tielens}, {Hartogh}, {St{\"u}tzki}, \& {Szczerba}}]{bergin2010}
{Bergin}, E.~A., {Hogerheijde}, M.~R., {Brinch}, C., {et~al.} 2010, \aap, 521,
  L33

\bibitem[{{Bolton} \& {Bolton}(2010)}]{bolton2010}
{Bolton}, S.~J., \& {Bolton}. 2010, in IAU Symposium, Vol. 269, IAU Symposium,
  ed. C.~{Barbieri}, S.~{Chakrabarti}, M.~{Coradini}, \& M.~{Lazzarin}, 92--100

\bibitem[{{Brasser} {et~al.}(2006){Brasser}, {Duncan}, \&
  {Levison}}]{brasser2006}
{Brasser}, R., {Duncan}, M.~J., \& {Levison}, H.~F. 2006, \icarus, 184, 59

\bibitem[Casassus et al.(2013)]{2013Natur.493..191C} 
 Casassus, S., van der Plas, G., M, S.~P., et al.\ 2013, \nat, 493, 191 

\bibitem[{{Ceccarelli} {et~al.}(2005){Ceccarelli}, {Dominik}, {Caux},
  {Lefloch}, \& {Caselli}}]{ceccarelli2005}
{Ceccarelli}, C., {Dominik}, C., {Caux}, E., {Lefloch}, B., \& {Caselli}, P.
  2005, \apjl, 631, L81

\bibitem[{{Chiang} \& {Goldreich}(1997)}]{chiang1997}
{Chiang}, E.~I., \& {Goldreich}, P. 1997, \apj, 490, 368

\bibitem[{{Ciesla}(2014)}]{ciesla2014}
{Ciesla}, F.~J. 2014, \apjl, 784, L1

\bibitem[{{Cochran}(2002)}]{cochran2002}
{Cochran}, A.~L. 2002, \apjl, 576, L165

\bibitem[{{Cochran} {et~al.}(2000){Cochran}, {Cochran}, \&
  {Barker}}]{cochran2000}
{Cochran}, A.~L., {Cochran}, W.~D., \& {Barker}, E.~S. 2000, \icarus, 146, 583

\bibitem[D'Alessio et al.(2006)]{2006ApJ...638..314D} 
 D'Alessio, P., Calvet, N., Hartmann, L., Franco-Hern{\'a}ndez, R., \& Serv{\'{\i}}n, H.\ 2006, \apj, 638, 314 

\bibitem[{{D'Angelo} \& {Marzari}(2012)}]{dangelo2012}
{D'Angelo}, G., \& {Marzari}, F. 2012, \apj, 757, 50

\bibitem[{{Desch}(2007)}]{desch2007}
{Desch}, S.~J. 2007, \apj, 671, 878

\bibitem[{{Dominik} {et~al.}(2005){Dominik}, {Ceccarelli}, {Hollenbach}, \&
  {Kaufman}}]{dominik2005}
{Dominik}, C., {Ceccarelli}, C., {Hollenbach}, D., \& {Kaufman}, M. 2005,
  \apjl, 635, L85

\bibitem[{{Dubrulle} {et~al.}(1995){Dubrulle}, {Morfill}, \&
  {Sterzik}}]{dubrulle1995}
{Dubrulle}, B., {Morfill}, G., \& {Sterzik}, M. 1995, \icarus, 114, 237

\bibitem[Espaillat et al.(2012)]{2012ApJ...747..103E} 
 Espaillat, C., Ingleby, L., Hern{\'a}ndez, J., et al.\ 2012, \apj, 747, 103 

\bibitem[{{Fatuzzo} \& {Adams}(2008)}]{fatuzzo2008}
{Fatuzzo}, M., \& {Adams}, F.~C. 2008, \apj, 675, 1361

\bibitem[{{Gautier} \& {Hersant}(2005)}]{gautier2005}
{Gautier}, D., \& {Hersant}, F. 2005, \ssr, 116, 25

\bibitem[{{Gautier} {et~al.}(2001){Gautier}, {Hersant}, {Mousis}, \&
  {Lunine}}]{gautier2001}
{Gautier}, D., {Hersant}, F., {Mousis}, O., \& {Lunine}, J.~I. 2001, \apjl,
  550, L227

\bibitem[{{Givan} {et~al.}(1996){Givan}, {Loewenschuss}, \&
  {Nielsen}}]{givan1996}
{Givan}, A., {Loewenschuss}, A., \& {Nielsen}, C.~J. 1996, Vib.~Spectrosc.,
  vol.~12, p.~1-14 (1996)., 12, 1

\bibitem[{{Gomes} {et~al.}(2005){Gomes}, {Levison}, {Tsiganis}, \&
  {Morbidelli}}]{gomes2005}
{Gomes}, R., {Levison}, H.~F., {Tsiganis}, K., \& {Morbidelli}, A. 2005, \nat,
  435, 466

\bibitem[{{Guillot} \& {Hueso}(2006)}]{guillot2006}
{Guillot}, T., \& {Hueso}, R. 2006, \mnras, 367, L47

\bibitem[{{Habing}(1968)}]{habing1968}
{Habing}, H.~J. 1968, \bain, 19, 421

\bibitem[{{Hersant} {et~al.}(2004){Hersant}, {Gautier}, \&
  {Lunine}}]{hersant2004}
{Hersant}, F., {Gautier}, D., \& {Lunine}, J.~I. 2004, \planss, 52, 623

\bibitem[{{Hogerheijde} {et~al.}(2011){Hogerheijde}, {Bergin}, {Brinch},
  {Cleeves}, {Fogel}, {Blake}, {Dominik}, {Lis}, {Melnick}, {Neufeld},
  {Pani{\'c}}, {Pearson}, {Kristensen}, {Y{\i}ld{\i}z}, \& {van
  Dishoeck}}]{hogerheijde2011}
{Hogerheijde}, M.~R., {Bergin}, E.~A., {Brinch}, C., {et~al.} 2011, Science,
  334, 338

\bibitem[{{Hollenbach} {et~al.}(1994){Hollenbach}, {Johnstone}, {Lizano}, \&
  {Shu}}]{hollenbach1994}
{Hollenbach}, D., {Johnstone}, D., {Lizano}, S., \& {Shu}, F. 1994, \apj, 428,
  654

\bibitem[{{Hudson} \& {Donn}(1991)}]{hudson1991}
{Hudson}, R.~L., \& {Donn}, B. 1991, \icarus, 94, 326

\bibitem[{{Irwin} {et~al.}(1998){Irwin}, {Weir}, {Smith}, {Taylor}, {Lambert},
  {Calcutt}, {Cameron-Smith}, {Carlson}, {Baines}, {Orton}, {Drossart},
  {Encrenaz}, \& {Roos-Serote}}]{irwin1998}
{Irwin}, P.~G.~J., {Weir}, A.~L., {Smith}, S.~E., {et~al.} 1998, \jgr, 103,
  23001

\bibitem[{{Johnstone} {et~al.}(1998){Johnstone}, {Hollenbach}, \&
  {Bally}}]{johnstone1998}
{Johnstone}, D., {Hollenbach}, D., \& {Bally}, J. 1998, \apj, 499, 758

\bibitem[{{Karssemeijer} {et~al.}(2014){Karssemeijer}, {de Wijs}, \&
  {Cuppen}}]{karssemeijer2014}
{Karssemeijer}, L.~J., {de Wijs}, G.~A., \& {Cuppen}, H.~M. 2014, {Phys. Chem.
  Chem. Phys.}, doi:10.1039/C4CP01622J

\bibitem[{{Kenyon} \& {Bromley}(2004)}]{kenyon2004}
{Kenyon}, S.~J., \& {Bromley}, B.~C. 2004, \nat, 432, 598

\bibitem[Kim et al.(2013)]{2013ApJ...769..149K} 
 Kim, K.~H., Watson, D.~M., Manoj, P., et al.\ 2013, \apj, 769, 149 

\bibitem[{{Kouchi}(1990)}]{kouchi1990}
{Kouchi}, A. 1990, Journal of Crystal Growth, 99, 1220

\bibitem[{{Kouchi} \& {Kuroda}(1990)}]{kouchi1990b}
{Kouchi}, A., \& {Kuroda}, T. 1990, in ESA Special Publication, Vol. 315, ESA
  Special Publication, ed. B.~{Battrick}, 193--196

\bibitem[{{Lada} \& {Lada}(2003)}]{lada2003}
{Lada}, C.~J., \& {Lada}, E.~A. 2003, \araa, 41, 57

\bibitem[{{Lesniak} \& {Desch}(2011)}]{lesniak2011}
{Lesniak}, M.~V., \& {Desch}, S.~J. 2011, \apj, 740, 118

\bibitem[{{Lissauer} {et~al.}(2009){Lissauer}, {Hubickyj}, {D'Angelo}, \&
  {Bodenheimer}}]{lissauer2009}
{Lissauer}, J.~J., {Hubickyj}, O., {D'Angelo}, G., \& {Bodenheimer}, P. 2009,
  \icarus, 199, 338

\bibitem[{{Lodders}(2003)}]{lodders2003}
{Lodders}, K. 2003, \apj, 591, 1220

\bibitem[{{Lyons} \& {Young}(2005)}]{lyons2005}
{Lyons}, J.~R., \& {Young}, E.~D. 2005, \nat, 435, 317

\bibitem[{{Mahaffy} {et~al.}(2000){Mahaffy}, {Niemann}, {Alpert}, {Atreya},
  {Demick}, {Donahue}, {Harpold}, \& {Owen}}]{mahaffy2000}
{Mahaffy}, P.~R., {Niemann}, H.~B., {Alpert}, A., {et~al.} 2000, \jgr, 105,
  15061

\bibitem[{{Mathis} {et~al.}(1977){Mathis}, {Rumpl}, \& {Nordsieck}}]{mrn77}
{Mathis}, J.~S., {Rumpl}, W., \& {Nordsieck}, K.~H. 1977, \apj, 217, 425

\bibitem[{{Mitchell} \& {Stewart}(2010)}]{mitchell2010}
{Mitchell}, T.~R., \& {Stewart}, G.~R. 2010, \apj, 722, 1115

\bibitem[{{Morbidelli} \& {Levison}(2004)}]{morbidelli2004}
{Morbidelli}, A., \& {Levison}, H.~F. 2004, \aj, 128, 2564

\bibitem[Najita et al.(2007)]{2007MNRAS.378..369N} 
 Najita, J.~R., Strom, S.~E., \& Muzerolle, J.\ 2007, \mnras, 378, 369 

\bibitem[{{Niemann} {et~al.}(1998){Niemann}, {Atreya}, {Carignan}, {Donahue},
  {Haberman}, {Harpold}, {Hartle}, {Hunten}, {Kasprzak}, {Mahaffy}, {Owen}, \&
  {Way}}]{niemann1998}
{Niemann}, H.~B., {Atreya}, S.~K., {Carignan}, G.~R., {et~al.} 1998, \jgr, 103,
  22831

\bibitem[{{Notesco} \& {Bar-Nun}(1996)}]{notesco1996}
{Notesco}, G., \& {Bar-Nun}, A. 1996, \icarus, 122, 118

\bibitem[{{{\"O}berg} {et~al.}(2009){{\"O}berg}, {Linnartz}, {Visser}, \& {van
  Dishoeck}}]{oberg2009}
{{\"O}berg}, K.~I., {Linnartz}, H., {Visser}, R., \& {van Dishoeck}, E.~F.
  2009, \apj, 693, 1209

\bibitem[{{Ouellette} {et~al.}(2007){Ouellette}, {Desch}, \&
  {Hester}}]{ouellette2007}
{Ouellette}, N., {Desch}, S.~J., \& {Hester}, J.~J. 2007, \apj, 662, 1268

\bibitem[{{Ouellette} {et~al.}(2010){Ouellette}, {Desch}, \&
  {Hester}}]{ouellette2010}
---. 2010, \apj, 711, 597

\bibitem[{{Owen} {et~al.}(1992){Owen}, {Bar-Nun}, \& {Kleinfeld}}]{owen1992}
{Owen}, T., {Bar-Nun}, A., \& {Kleinfeld}, I. 1992, \nat, 358, 43

\bibitem[{{Owen} \& {Encrenaz}(2003)}]{owen2003}
{Owen}, T., \& {Encrenaz}, T. 2003, \ssr, 106, 121

\bibitem[{{Owen} \& {Encrenaz}(2006)}]{owen2006}
---. 2006, \planss, 54, 1188

\bibitem[{{Owen} {et~al.}(1999){Owen}, {Mahaffy}, {Niemann}, {Atreya},
  {Donahue}, {Bar-Nun}, \& {de Pater}}]{owen1999}
{Owen}, T., {Mahaffy}, P., {Niemann}, H.~B., {et~al.} 1999, \nat, 402, 269

\bibitem[{{Pan} {et~al.}(2012){Pan}, {Desch}, {Scannapieco}, \&
  {Timmes}}]{pan2012}
{Pan}, L., {Desch}, S.~J., {Scannapieco}, E., \& {Timmes}, F.~X. 2012, \apj,
  756, 102

\bibitem[{{Podio} {et~al.}(2013){Podio}, {Kamp}, {Codella}, {Cabrit}, {Nisini},
  {Dougados}, {Sandell}, {Williams}, {Testi}, {Thi}, {Woitke}, {Meijerink},
  {Spaans}, {Aresu}, {M{\'e}nard}, \& {Pinte}}]{podio2013}
{Podio}, L., {Kamp}, I., {Codella}, C., {et~al.} 2013, \apjl, 766, L5

\bibitem[Rosenfeld et al.(2014)]{2014ApJ...782...62R} 
 Rosenfeld, K.~A., Chiang, E., \& Andrews, S.~M.\ 2014, \apj, 782, 62 

\bibitem[{{Roulston} \& {Stevenson}(1995)}]{roulston1995}
{Roulston}, M.~S., \& {Stevenson}, D.~J. 1995, {EOS}, 76, 343

\bibitem[{{Sano} {et~al.}(2000){Sano}, {Miyama}, {Umebayashi}, \&
  {Nakano}}]{sano2000}
{Sano}, T., {Miyama}, S.~M., {Umebayashi}, T., \& {Nakano}, T. 2000, \apj, 543,
  486

\bibitem[{{Stevenson} \& {Salpeter}(1977{\natexlab{a}})}]{stevenson1977b}
{Stevenson}, D.~J., \& {Salpeter}, E.~E. 1977{\natexlab{a}}, \apjs, 35, 239

\bibitem[{{Stevenson} \& {Salpeter}(1977{\natexlab{b}})}]{stevenson1977a}
---. 1977{\natexlab{b}}, \apjs, 35, 221

\bibitem[{{Wadhwa} {et~al.}(2007){Wadhwa}, {Amelin}, {Davis}, {Lugmair},
  {Meyer}, {Gounelle}, \& {Desch}}]{wadhwa2007}
{Wadhwa}, M., {Amelin}, Y., {Davis}, A.~M., {et~al.} 2007, Protostars and
  Planets V, 835

\bibitem[{{Walsh} {et~al.}(2011){Walsh}, {Morbidelli}, {Raymond}, {O'Brien}, \&
  {Mandell}}]{walsh2011}
{Walsh}, K.~J., {Morbidelli}, A., {Raymond}, S.~N., {O'Brien}, D.~P., \&
  {Mandell}, A.~M. 2011, \nat, 475, 206


\bibitem[{{Weaver} {et~al.}(2002){Weaver}, {Feldman}, {Combi}, {Krasnopolsky},
  {Lisse}, \& {Shemansky}}]{weaver2002}
{Weaver}, H.~A., {Feldman}, P.~D., {Combi}, M.~R., {et~al.} 2002, \apjl, 576,
  L95

\bibitem[{{Weidenschilling}(1977{\natexlab{a}})}]{weidenschilling1977b}
{Weidenschilling}, S.~J. 1977{\natexlab{a}}, \mnras, 180, 57

\bibitem[{{Weidenschilling}(1977{\natexlab{b}})}]{weidenschilling1977a}
---. 1977{\natexlab{b}}, \apss, 51, 153

\bibitem[{{Weidenschilling}(1997)}]{weidenschilling1997}
---. 1997, \icarus, 127, 290

\bibitem[{{Westley} {et~al.}(1995){Westley}, {Baragiola}, {Johnson}, \&
  {Baratta}}]{westley1995}
{Westley}, M.~S., {Baragiola}, R.~A., {Johnson}, R.~E., \& {Baratta}, G.~A.
  1995, \nat, 373, 405

\bibitem[{{Wilson} \& {Militzer}(2010)}]{wilson2010}
{Wilson}, H.~F., \& {Militzer}, B. 2010, Physical Review Letters, 104, 121101

\bibitem[{{Woitke} {et~al.}(2009){Woitke}, {Thi}, {Kamp}, \&
  {Hogerheijde}}]{woitke2009}
{Woitke}, P., {Thi}, W.-F., {Kamp}, I., \& {Hogerheijde}, M.~R. 2009, \aap,
  501, L5

\bibitem[{{Wong} {et~al.}(2004){Wong}, {Mahaffy}, {Atreya}, {Niemann}, \&
  {Owen}}]{wong2004}
{Wong}, M.~H., {Mahaffy}, P.~R., {Atreya}, S.~K., {Niemann}, H.~B., \& {Owen},
  T.~C. 2004, \icarus, 171, 153

\bibitem[{{Young} {et~al.}(2014){Young}, {Desch}, {Anbar}, {Barnes}, {Hinkel},
  {Kopparappu}, {Madhusudhan}, {Monga}, {Pagano}, {Riner}, {Scannapieco},
  {Shim}, \& {Truitt}}]{young2014}
{Young}, P.~A., {Desch}, S.~J., {Anbar}, A.~D., {et~al.} 2014, Astrobiology, in
  press

\end{thebibliography}
\end{document}